\documentclass[11pt,a4paper]{article}

\usepackage[T1]{fontenc}
\usepackage[margin=25mm]{geometry}
\usepackage{amsmath,amssymb}
\usepackage{bm}
\usepackage{graphicx}
\usepackage{booktabs}
\usepackage[section]{placeins}
\usepackage[numbers,sort&compress]{natbib}
\usepackage[hidelinks]{hyperref}

\DeclareRobustCommand{\quantity}[2]{\ensuremath{#1\,\mathrm{#2}}}
\renewcommand{\Re}{\mathfrak{Re}}
\title{Transverse Modulation of Continuous Electron Beams by a Structured Optical Cavity}
\author{%
  Marius Constantin Chirita Mihaila\textsuperscript{*},
  Julie Sk{\'y}palov{\'a} and Martin Koz{\'a}k\textsuperscript{*}\\[0.6em]
  \normalsize Charles University, Faculty of Mathematics and Physics,\\
  \normalsize Ke Karlovu 3, 121 16 Prague 2, Czech Republic\\[0.3em]
  \small \textsuperscript{*}Corresponding authors\quad
  \href{mailto:marius.chirita@matfyz.cuni.cz}{marius.chirita@matfyz.cuni.cz}\quad
  \href{mailto:m.kozak@matfyz.cuni.cz}{m.kozak@matfyz.cuni.cz}%
}
\date{}
\hypersetup{
  pdftitle={Transverse Modulation of Continuous Electron Beams by a Structured Optical Cavity},
  pdfauthor={Marius Constantin Chirita Mihaila, Julie Sk{\'y}palov{\'a}, Martin Koz{\'a}k}
}

\begin{document}
\maketitle

\begin{abstract}
Compact aberration correction at large semi-angles remains a central challenge in electron microscopy. Here we propose a phase plate for continuous electron beams based on the ponderomotive interaction with an intracavity standing wave in a near-concentric Fabry--P\'erot resonator. For electrons propagating along the cavity axis, a standing wave with a Laguerre--Gaussian mode of topological charge $\ell=1$ imprints an annular phase shift opposite in sign to the third-order spherical aberration of a conventional electron lens. For an electron beam with a mean energy
of 5~keV, a convergence semi-angle of 25~mrad,
and a Gaussian energy-distribution FWHM of 0.1~eV,
we predict that optical spherical-aberration correction
of an objective lens with $C_s=C_c=1\,\mathrm{mm}$ produces
a probe with a diameter enclosing 50\% of the electron
current of approximately 0.46~nm.
The energy-averaged probe has an effective Strehl ratio
of approximately $0.77$, defined as its peak intensity
relative to that of the monochromatic diffraction-limited
probe for the same pupil and total electron current. The accessible transverse electron phases are determined by the resonator-supported modal basis.
\end{abstract}


\section{Introduction}\label{sec:level1}

Low-energy electron microscopy, spanning beam energies from single-digit \ensuremath{\mathrm{eV}} to \quantity{30}{keV}, offers an attractive route to high-resolution imaging and spectroscopy while mitigating radiation damage relative to conventional TEM operation \cite{jobst2016quantifying,geelen2019nonuniversal,sunaoshi2016stem,chirita2022three}. Recent work has further shown that electron energy-loss spectroscopy can access rich quantum and optical excitations at energies well below \quantity{30}{keV}~\cite{shiloh2022quantum,simonaitis2025electron}, while advances in monochromation are pushing energy spreads into the single-digit-\ensuremath{\mathrm{meV}} regime \cite{lagos2022advances}. However, achieving atomic-resolution spectroscopy under these conditions requires aberration correction at large semi-angles, which remains challenging in compact low-kV instruments.

In parallel, structured electron beams have emerged as a versatile tool for tailoring interaction and contrast, with demonstrated control over transverse phase and amplitude using holographic and diffractive approaches \cite{verbeeck2010production,mcmorran2011electron,grillo2014}. For high-resolution imaging, dedicated multipole correctors can compensate for primary aberrations and enable atomic resolution in advanced columns \cite{haider1998electron}, and recent efforts explore alternative implementations such as wire-based correctors \cite{nakano2025development}. Although powerful, these systems are complex and bulky, motivating complementary approaches that provide tunable, compact wavefront control.

Electron-wave shaping by light has been demonstrated near optically excited thin films and nanostructures, where optical near fields imprint tailored spatiotemporal structure on the beam \cite{barwick2009photon,feist2015quantum,vanacore2019ultrafast,konevcna2020electron,shiloh2021electron,henke2021integrated,feist2022cavity,madan2022ultrafast,tsesses2023tunable,garcia2023spatiotemporal,fang2024structured,ferrari2025realization, garcia2025roadmap,chahshouri2026stimulated}. For aberration compensation, tailored corrections can be implemented with microfabricated phase elements, including sculpted thin-film phase plates~\cite{shiloh2018spherical} and pixelated electrostatic devices \cite{Tavabi2021,vega2023can,adelmark2026fabrication}. However, material implementations are lossy and susceptible to charging, contamination, and dose-induced degradation, while electrostatic approaches introduce their own challenges in fabrication, fringing-field control, and diffraction at sub-apertures. These limitations have motivated a growing interest in light-based electron optics in free space \cite{freimund2001observation,hebeisen2006femtosecond,de2021optical,chirita2022transverse,chirita2025design,mihaila2025light,chirita2025light,uesugi2021electron,streshkova2024monochromatization,uesugi2022properties,uesugi2025crossed,nekula2025compensating,uesugi2026achromat,ebel2025structured,chahshouri2026stimulated}, where the electron phase can be modulated without the need to pass through a material phase plate.

Continuous-wave implementations are particularly appealing for steady-state sources, spanning single-pass geometries \cite{velasco2025free,zhao2025gouy} and cavity-enhanced schemes that reach practical phase shifts for beam shaping \cite{henke2021integrated,feist2022cavity}. Complementary mechanisms have also been proposed in which the electron beam acts as an effective refractive medium for photons \cite{zheng2026elastic}.

Nevertheless, imprinting substantial phase shifts on continuous electron waves with continuous-wave light in free space requires cavity enhancement to reach the required optical intensities. Near-concentric Fabry--P\'erot resonators can be designed for tight focusing, and such operation has been demonstrated from low circulating powers \cite{durak2014diffraction,nguyen2018operating,utama2021coupling} to the $>\!100~\mathrm{kW}$ regime \cite{schwartz2017near,schwartz2019laser,PhysRevLett.124.174801,petrov2026crossed,Petrov2026LaserPhasePlate,arakcheev2026probing}. Indeed, cavity implementations have achieved continuous-wave intracavity intensities above $450~\mathrm{GW/cm^2}$, producing a $\pi/2$ phase shift for $300~\mathrm{keV}$ electrons over a nominal transverse interaction length of approximately $ 17~\mu\mathrm{m}$ in phase-contrast electron microscopy \cite{turnbaugh2021high}.

Recent experiments established optical correction of third-order electron spherical aberration using a shaped pulsed light field based on a charge-1 Laguerre--Gaussian beam (LG$_0^1$)~\cite{chirita2025light}. Here we extend that approach to a cavity-enhanced continuous-wave architecture for continuous electron beams. On--axis electron-field coupling maximizes the usable interaction length, enabling large accumulated phase shifts while preserving cylindrical symmetry. The attainable transverse electron phase profiles are constrained by the resonantly excited optical eigenmodes supported by the cavity, which may be preferentially addressed by shaping the incident field, for example with a spatial light modulator (SLM)~\cite{zhu2014arbitrary,utama2021coupling} or a specially designed phase plate based on metaoptics~\cite{beresna2011radially}.

This mode-limited yet reconfigurable platform bridges adaptive electron optics and resonant photonics, enabling phase engineering of continuous electron beams at experimentally accessible power levels. As we show below, the perforated cavity can still sustain an intracavity LG$_0^1$-like mode, supporting the feasibility of aberration correction toward atomic resolution at electron energies relevant to widely used scanning electron microscopes.

\section{Ponderomotive phase model for a structured intracavity field}\label{sec:level2}

We consider a monochromatic electron of wavelength $\lambda_\mathrm{e}$
traversing the intracavity optical field and acquiring a ponderomotive phase
shift. In the projected-potential approximation, the phase imprinted on the
electron wave function is~\cite{de2021optical}
\begin{equation}
    \phi_\mathrm{p}(\bm r_\perp)
    =
    -\frac{1}{\hbar}
    \int_{t_0}^{t_1}\mathrm{d}t\,U[\bm r(t),t],
    \label{eq:phase_finite_time}
\end{equation}
where $\bm r(t)=(x,y,z(t))$ denotes the trajectory of the electron through the optical
field and $\hbar$ is the reduced Planck constant. The transverse impact coordinate $\bm r_\perp=(x,y)$ is taken to be
constant during the interaction, and the longitudinal motion is approximated as
uniform $z(t)=z_0+vt$ (or, by a suitable choice of origin, $z(t)=vt$) where \(z_0\) is the longitudinal position of the electron at \(t=0\). The integration interval
$(t_0,t_1)$ spans the full electron--light interaction. These approximations are applicable in the case of collimated electron beam and a weak electron-photon interaction. The
ponderomotive potential is
\begin{equation}
U(\bm r,t)
=
\frac{e^2}{2m_e\gamma}
\left[
A_x^2(\bm r,t)
+
A_y^2(\bm r,t)
+
\frac{A_z^2(\bm r,t)}{\gamma^2}
\right],
\label{eq:Ponderomotive pontential original}
\end{equation}
where $e$ is the elementary charge, $m_e$ is the electron mass, $c$ is the speed of light in vacuum, $\gamma=1/\sqrt{1-v^2/c^2}$ is the relativistic factor, and $v$ is the electron velocity.

The corresponding physical vector potential is
\begin{equation}
\bm A(\bm r,t)
=
\Re\!\left\{
\frac{\bm E(\bm r,t)}{i\omega_0}
\right\},
\label{eq:vector_potential}
\end{equation}
where $\bm E(\bm r,t)$ denotes the complex representation of the
monochromatic intracavity electric field, with time dependence
$e^{-i\omega_0 t}$, and $\omega_0$ is its angular frequency. The intracavity standing wave is formed by two
counter-propagating Laguerre--Gaussian beams with radial index $p$
and azimuthal index $\ell$. The full electric-field expression,
including the longitudinal component, is given in
Appendix~\ref{app:field}, Eq.~(\ref{eq:LG mode of the cavity}).
Its spatial dependence is written in cylindrical coordinates
$(\rho,\varphi,z)$, with $x=\rho\cos\varphi$ and
$y=\rho\sin\varphi$, while the field components are referred to
the fixed Cartesian axes. This field determines the ponderomotive
potential and hence the electron phase through the relations above.

\begin{figure}[t]
    \centering
    \includegraphics[width=10cm]{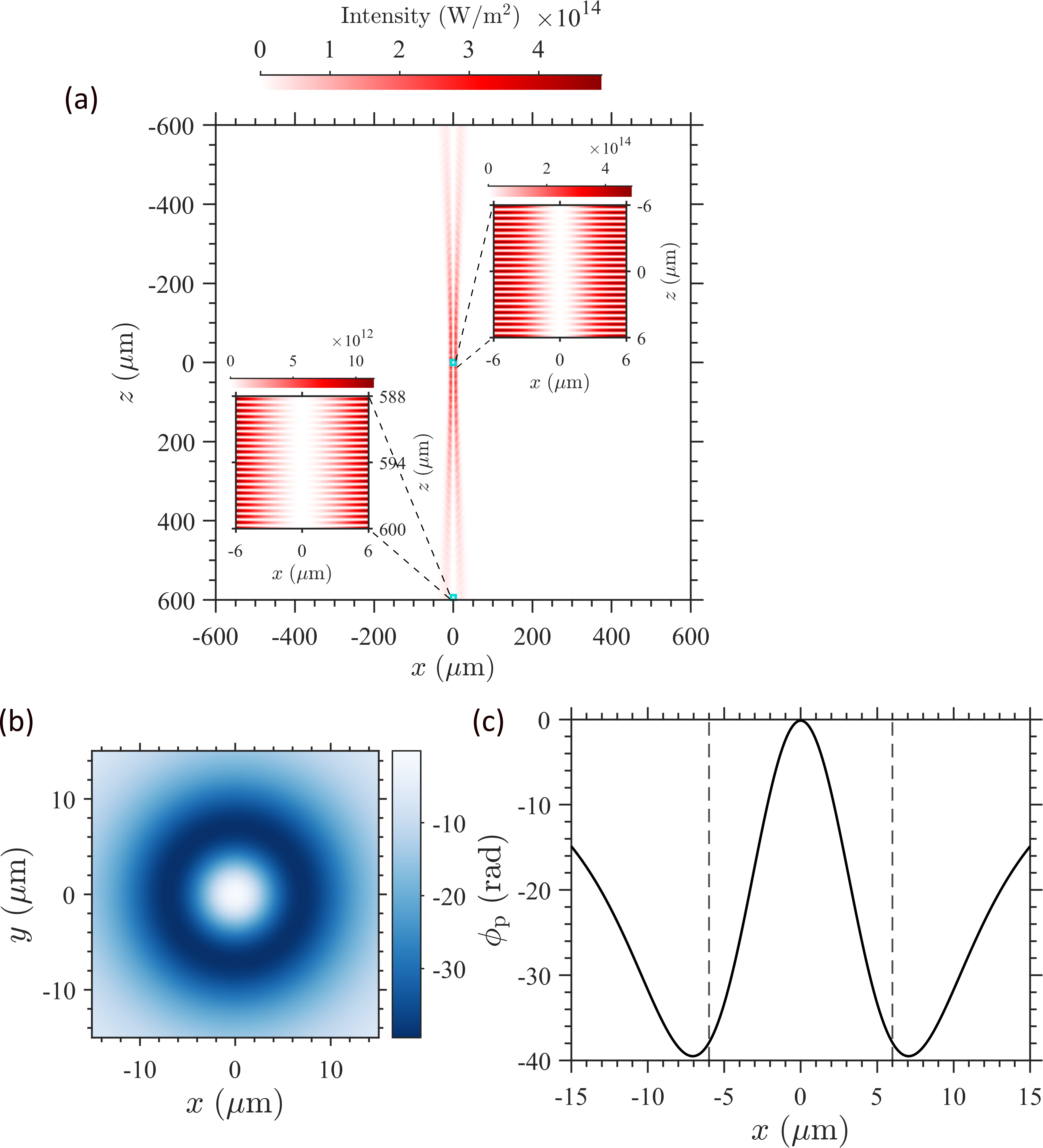}
    \caption{(a) Standing-wave intensity distribution of the optical
    $\mathrm{LG}_{0}^{1}$ mode in the $y=0$ plane (longitudinal $x$--$z$ cut),
    for $\lambda_\mathrm{L}=\quantity{1064}{nm}$ and laser waist
    $w_\mathrm{f}=\quantity{8.1}{\text{\textmu}m}$ (calculated using Eq.~\ref{eq:LG mode of the cavity} in appendix~\ref{app:field}). Insets show magnified views of the
    axial standing-wave modulation at the two marked longitudinal positions.
    The intensity decreases by approximately one and a half orders of magnitude at a
    longitudinal distance of \quantity{600}{\text{\textmu}m} from the focal spot.
    (b) Transverse electron phase profile $\phi_\mathrm{p}(x,y)$ accumulated by an
    electron traversing the field along $z$. Color scales indicate intensity in
    (a) and ponderomotive phase shift in (b). (c) Line profile of the phase shift
    from panel (b) taken along the $x$ axis at $y=0$. The dashed vertical lines at
    $x=\pm\quantity{6}{\text{\textmu}m}$ mark the transverse region of the optical standing wave
    sampled by the electron beam, showing that the electrons acquire phase only from
    the central part of the field within $|x|\leq\quantity{6}{\text{\textmu}m}$.}
    \label{fig:I donut and phase}
\end{figure}

In the calculations below, the interaction is treated as a thin phase element,
so that the electron exit wave is
\begin{equation}
\psi_{\rm mod}(x,y,z=0)\simeq \psi_{\rm in}(x,y,z=0)\exp[i\phi_\mathrm{p}(x,y)],
\label{eq:thin_phase_exit}
\end{equation}
where $\psi_{\mathrm{in}}$ and $\psi_{\mathrm{mod}}$ are the electron wave functions before and after phase modulation, respectively, and $\phi_p(x,y)$ is obtained from Eq. (1) by integrating the ponderomotive potential along the electron trajectory through the optical field. As an illustrative
example, Fig.~\ref{fig:I donut and phase}(a) shows the standing-wave intensity
distribution of an $\mathrm{LG}_{0}^{1}$ intracavity mode, and
Fig.~\ref{fig:I donut and phase}(b) shows the corresponding accumulated
transverse electron phase profile. For the parameters used here, integration over the
approximate interaction length
$d_\mathrm{int}\simeq\quantity{1.2}{mm}$ yields a peak phase shift of
$\sim 39$~rad (see Fig~\ref{fig:I donut and phase}).

Internal free-space propagation within the interaction region, i.e., Fresnel- or multi-slice propagation during the electron--light interaction, is neglected in the present model. Assuming that the electron beam enters the cavity with the maximum convergence/divergence semi-angle of $|\theta_\mathrm{IP}|\lesssim 0.5~\mathrm{mrad}$, the geometric broadening over the interaction length is only
$\Delta r_\mathrm{geom}\simeq d_\mathrm{int}\theta_\mathrm{IP}=0.6~\mu\mathrm{m}$, which is much less than the characteristic dimensions of the cavity mode used to induce the phase modulation of the electrons.
Any additional transverse deflection induced by the optical phase modulation is still smaller and is generated predominantly near the cavity focus, where the optical intensity is highest [see Fig.~1(a)]. Hence, estimating the corresponding transverse displacement using the full interaction length $d_\mathrm{int}$ is conservative. Transverse redistribution during the interaction is therefore negligible, which justifies the projected-potential approximation.

\section{Results: Third-order spherical-aberration correction}\label{sec:level3}
\begin{figure}[t]
    \centering
    \includegraphics[width=10cm]{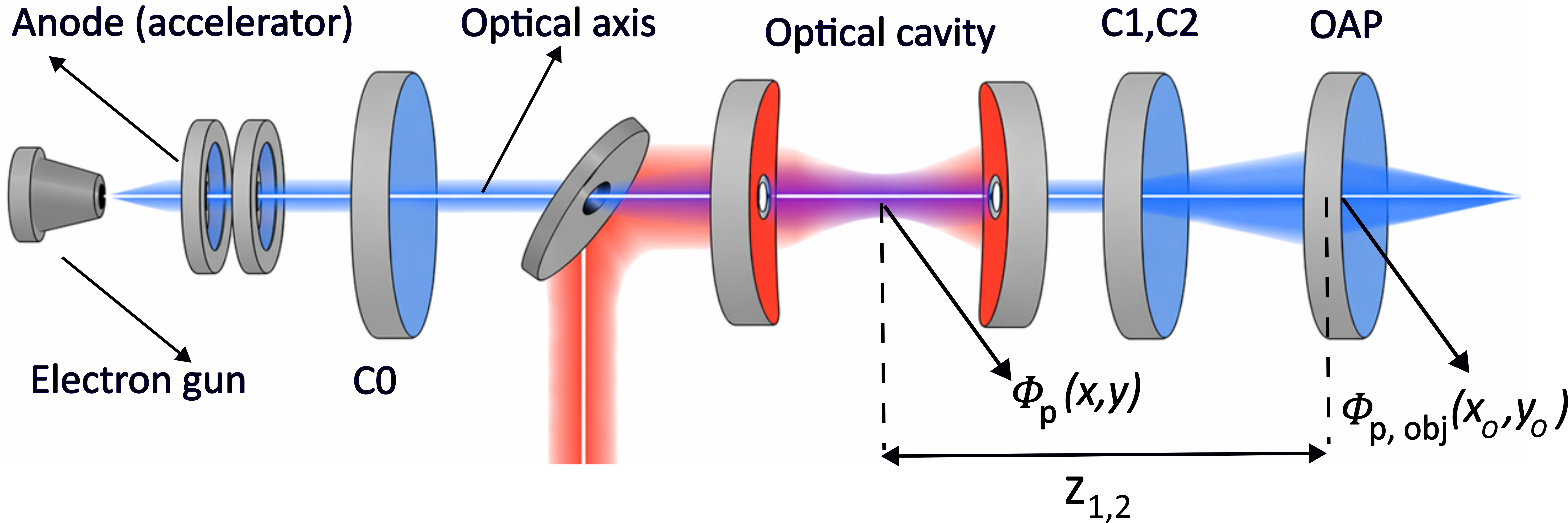}
    \caption{\label{fig:fig2} Proof-of-concept electron-optical layout for transverse phase modulation of a continuous electron beam by a structured mode of an optical cavity. An electron beam is prepared by a condenser lens/lenses (C0) and enters the cavity, where the circulating light imprints a ponderomotive phase $\phi_{\mathrm{p}}(x,y)$ at the interaction plane (IP). The IP is then imaged by a second condenser system (C1, C2) onto its conjugate plane at the objective aperture plane (OAP) at an axial distance $z_{1,2}$ from the IP, with magnification $M$, yielding the phase profile $\phi_{\mathrm{p,obj}}(x_{\mathrm{0}}, y_{\mathrm{0}})$. The light can be coupled into the cavity using one additional flat mirror oriented at $45^\circ$ with respect to the optical axis. The optical elements brought to the electron microscope column contain a small central aperture to allow the electron beam to pass through unobstructed.}
    \label{fig:Schematic Microscope}
\end{figure}

We now apply the phase-plate model above to the compensation of third-order
spherical aberration. For analytical simplicity, we assume that the
electron--light interaction plane (IP) is imaged onto the objective
aperture plane (OAP), or equivalently onto an aperture-conjugate pupil plane,
with transverse magnification $M$ (here $M \simeq 4.2$, mapping
$\sim 12~\mu\mathrm{m}$ features at the IP to $\sim 50~\mu\mathrm{m}$ features
in the OAP). Under this first-order imaging assumption, the ponderomotive
phase profile is transferred to the OAP as a transversely rescaled phase
distribution, up to a quadratic phase term that is absorbed into defocus.
Exact conjugate-plane alignment is not required in principle; however, in the more general case of a non-conjugate configuration, the phase at the OAP must be computed by explicit propagation through the intervening optics.

Let $(x,y)$ denote the transverse coordinates in the interaction plane and
$(x_\mathrm{0},y_\mathrm{0})$ those in the objective aperture plane. The two
coordinate systems are related by
\begin{equation}
x=\frac{x_\mathrm{0}}{M},
\qquad
y=\frac{y_\mathrm{0}}{M}.
\label{eq:coord_map}
\end{equation}
For a first-order (ABCD) imaging system operated under exact imaging conditions
($B=0$, where $B$ is the $(1,2)$ element of the ray-transfer matrix), the paraxial wave-optical mapping from IP to OAP is~\cite{fienup2024fourier}
\begin{equation}
\begin{aligned}
\psi_{\mathrm{OAP}}(x_\mathrm{0},y_\mathrm{0})
&=
\frac{e^{ik_eL_0}}{M}
\exp\!\left[
i\,\frac{k_e}{2}\,\frac{C}{M}
\left(x_\mathrm{0}^2+y_\mathrm{0}^2\right)
\right]
\\
&\quad\times
\psi_{\mathrm{mod}}\!\left(\frac{x_\mathrm{0}}{M},\frac{y_\mathrm{0}}{M}\right),
\end{aligned}
\label{eq:relay_psi}
\end{equation}
where $k_e=2\pi/\lambda_e$ is the electron wave number, $L_0$ is a path-length
constant contributing only a global phase, and $C$ is the $(2,1)$ element of
the ray-transfer matrix.

Consequently, a phase-only modulation at the interaction plane is transferred
to the objective plane as
\begin{equation}
\phi_{\mathrm{p},\mathrm{obj}}(x_\mathrm{0}, y_\mathrm{0})
=
\phi_\mathrm{p}\!\left(\frac{x_\mathrm{0}}{M}, \frac{y_\mathrm{0}}{M}\right)
\label{eq:relay_phase}
\end{equation}
that is, the phase values are preserved under the relay, while their transverse
scale is set by $M$. The additional quadratic phase in
Eq.~(\ref{eq:relay_psi}) is equivalent to a small effective defocus and can be
absorbed into the defocus of the objective lens. In what follows, we
omit the global phase factor $e^{ik_eL_0}$, the overall prefactor $1/M$, and
this relay-induced defocus term.

In the OAP, the electron wave function is
\begin{equation}
\begin{aligned}
\psi_{\mathrm{obj}}(x_\mathrm{0},y_\mathrm{0})
&=
A(x_\mathrm{0},y_\mathrm{0})
\\
&\quad\times
\exp\!\left\{
i\!\left[
\phi_{\mathrm{p},\mathrm{obj}}(x_\mathrm{0},y_\mathrm{0})
-
\chi\!\bigl(\theta(x_\mathrm{0},y_\mathrm{0})\bigr)
\right]
\right\},
\end{aligned}
\label{eq:pupil}
\end{equation}
where $A(x_\mathrm{0},y_\mathrm{0})$ is the pupil function and
$\theta(x_\mathrm{0},y_\mathrm{0})$ is the local convergence angle with respect
to the optic axis. The primary lens-aberration phase is~\cite{scherzer1947spharische}
\begin{equation}
\chi\!\bigl(\theta(x_\mathrm{0},y_\mathrm{0})\bigr)
=
\frac{\pi}{2\lambda_e}
\left[
C_s\,\theta^4(x_\mathrm{0},y_\mathrm{0})
-
2\,\Delta z\,\theta^2(x_\mathrm{0},y_\mathrm{0})
\right],
\label{eq:aberration}
\end{equation}
where $C_s$ is the third-order spherical-aberration coefficient and
$\Delta z$ is the defocus.

The probe in the Fourier plane is~\cite{haider2000upper}
\begin{equation}
\psi_f(x_f,y_f)=\mathcal{F}\{\psi_{\mathrm{obj}}(x_\mathrm{0},y_\mathrm{0})\},
\label{eq:probe_fft}
\end{equation}
where $(x_f,y_f)$ denote the spatial coordinates in the Fourier plane and \(\mathcal F\) denotes the two-dimensional Fourier transform. The corresponding radial coordinate in
the focal plane is $r_f=\sqrt{x_f^2+y_f^2}$. Its
intensity is
\begin{equation}
I(x_f,y_f)=\left|\psi_f(x_f,y_f)\right|^2.
\label{eq:intensity}
\end{equation}

\begin{figure}[t]
    \centering
    \includegraphics[width=10cm]{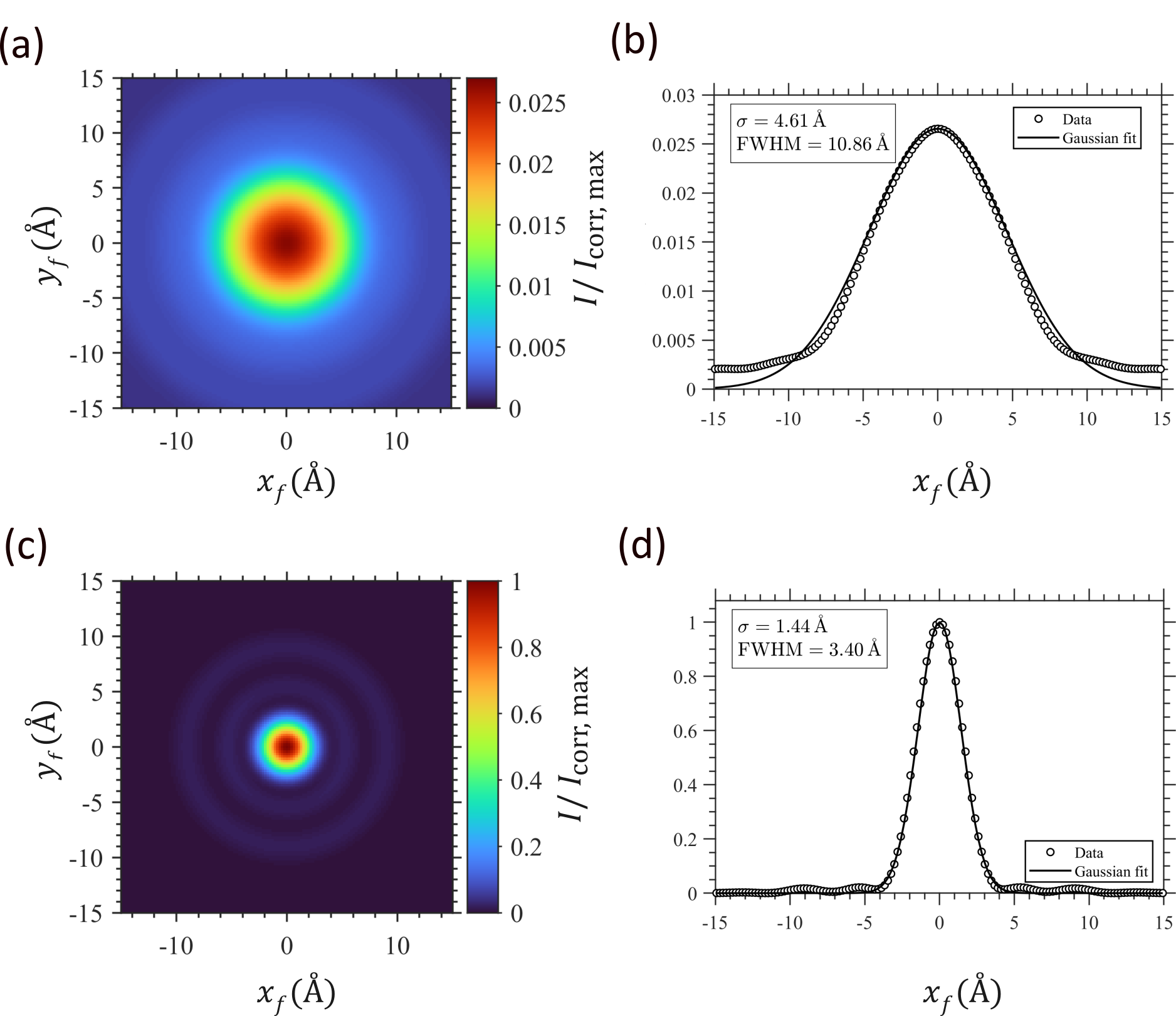}
    \caption{Monochromatic focal-plane electron-probe intensities before
and after ponderomotive wavefront correction. 
(a) Uncorrected probe intensity dominated by primary spherical aberration and (b) its central line profile with a Gaussian fit. 
(c) Corrected probe intensity and (d) its central line profile with a Gaussian fit. 
All intensities are normalized to the peak intensity of the corrected probe, $I_{\mathrm{corr,max}}$, such that the corrected peak is unity and the uncorrected peak directly gives the relative peak intensity. 
Primary spherical aberration reduces the central-lobe peak intensity to about $3\%$ of the corrected value. 
A different colorbar range is used in (a) to make the uncorrected focal spot visible on its respective scale.}
    \label{fig:focii}
\end{figure}

To demonstrate cavity-mediated wavefront shaping, we synthesize a compensating
phase
$\phi_{p,\mathrm{obj}}(x_\mathrm{0},y_\mathrm{0})-\chi\!\bigl(\theta(x_\mathrm{0},y_\mathrm{0})\bigr)= \phi_{\rm res}(x_\mathrm{0},y_\mathrm{0})\approx\mathrm{const}$
within the cavity-supported modal basis, such that the residual phase
$\phi_{\rm res}(x_\mathrm{0},y_\mathrm{0})$ is minimized across the objective-lens pupil,
as shown in appendix~\ref{app:pupil}, Fig.~\ref{fig:pupil-phase-maps}(c,d).

For an electron-lens focal length $f=\quantity{1}{mm}$, $C_\mathrm{s}=\quantity{1}{mm}$ and an
electron-beam radius at the OAP of $\rho_\mathrm{e}=\quantity{25}{\text{\textmu}m}$,
the corrected probe approaches the diffraction limit at a semi-convergence
angle of \quantity{25}{mrad} [Fig.~\ref{fig:focii}(c)]. The optimum compensation is
obtained for $\Delta z=\quantity{638}{nm}$, for which the corrected focal spot has
a fitted Gaussian width of $\sigma_\mathrm{p}\simeq\quantity{1.4}{\text{\AA}}$
[Fig.~\ref{fig:focii}(d)].

A Gaussian fit to the central line profiles
of the monochromatic probes
gives an apparent reduction of the probe width by approximately
a factor of three, from
$\sigma\simeq\quantity{4.6}{\text{\AA}}$ for the aberrated probe to
$\sigma\simeq\quantity{1.4}{\text{\AA}}$ after correction.
These fitted Gaussian widths characterize only the central lobes of the monochromatic probes and do not describe the full current distributions, which include surrounding rings and tails. They should be distinguished from FW50, the diameter enclosing 50$\%$ of the total electron current, reported below for the corrected probe.

\begin{figure}[htbp]
    \centering
    \includegraphics[width=12cm]
        {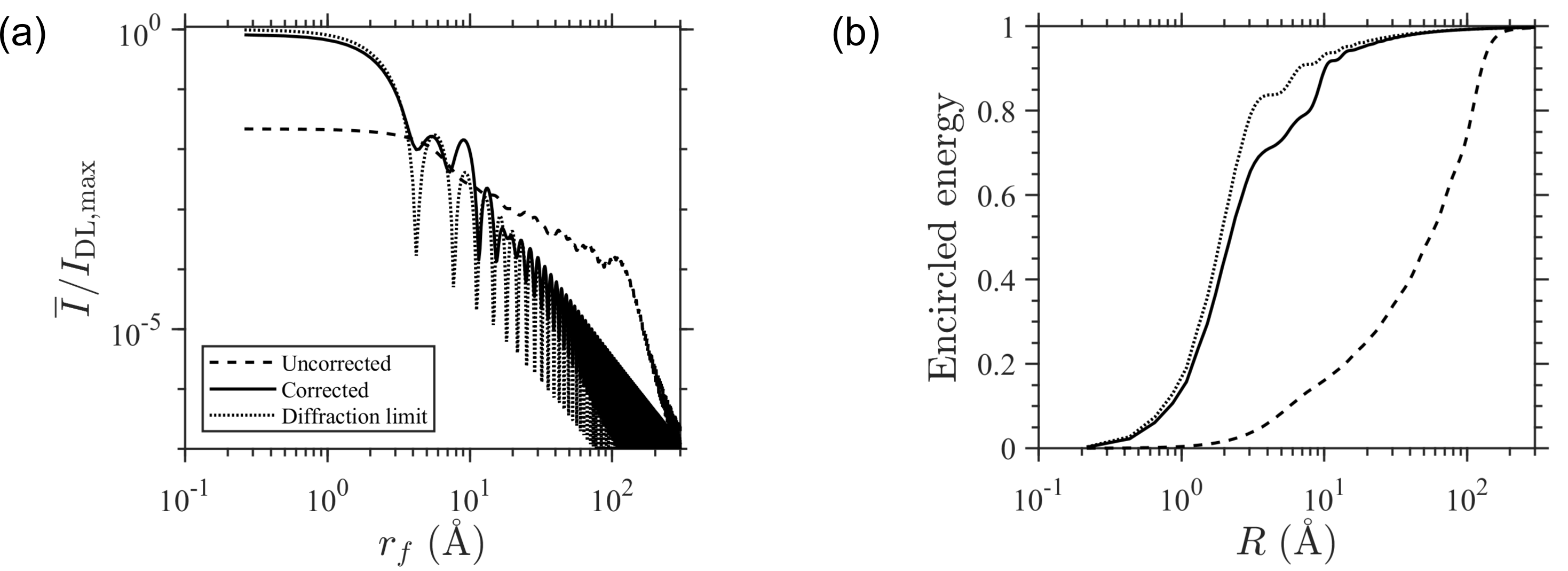}
    \caption{Monochromatic electron-probe quality before and after
ponderomotive correction for a $5\,\mathrm{keV}$ electron
beam and a convergence semi-angle of $25\,\mathrm{mrad}$.
    (a) Azimuthally averaged focal-plane intensity,
    normalized to the diffraction-limited peak, shown on
    logarithmic axes to reveal the surrounding rings and
    low-intensity tails.
    (b) Encircled-energy fraction: the total electron
    probability enclosed within a circle of radius $R$.
    Dashed, solid and dotted curves denote the uncorrected,
    corrected and diffraction-limited probes, respectively.
    All probes use the same uniform circular pupil and are
    normalized to equal total electron probability.
    The uncorrected and corrected probes are evaluated at
    defocus values of $0$ and $638\,\mathrm{nm}$, respectively.}
    
    \label{fig:probe_quality}
\end{figure}

Figure~\ref{fig:probe_quality} compares the monochromatic
uncorrected and $C_\mathrm{s}$-corrected probes with a diffraction-limited
reference to assess probe quality beyond the central
Gaussian width. All probes use the same uniform circular
pupil and electron wavelength and are normalized to equal
total electron probability integrated over the transverse
beam profile.

Figure~\ref{fig:probe_quality}(a) shows the azimuthally averaged focal-plane
intensity on logarithmic axes, normalized to the
diffraction-limited peak intensity $I_{\mathrm{DL},\max}$.
We define the Strehl-like peak-intensity ratio as
$S=I_{\max}/I_{\mathrm{DL},\max}$, where $I_{\max}$ denotes
the peak intensity of the calculated probe. This ratio
increases from $0.0219$ without correction to $0.826$ with ponderomotive
correction, quantifying the recovery of the central peak.

Figure~\ref{fig:probe_quality}(b) characterizes the spatial concentration of
electron probability through the encircled-energy fraction:
the fraction of the total electron probability contained
within a circle of radius $R$ centred on the probe, as
defined in Eq.~(B7) in
Appendix~\ref{subsec:finite_energy_probe}.
Ponderomotive correction increases this fraction at small
radii, demonstrating improved current concentration.
The corrected probe nevertheless retains more probability
in the surrounding rings than the diffraction-limited
reference, revealing the remaining departure from the
ideal probe. At the applied defocus of $638\,\mathrm{nm}$, the residual
phase shown in Fig.~B1(c) has a rms error
of $0.432\,\mathrm{rad}$ and a peak-to-valley variation
of $1.269\,\mathrm{rad}$ over the full pupil. The rms is
calculated after subtraction of the mean pupil phase;
no additional defocus term is fitted or removed. The maximum azimuthally averaged
intensity beyond the first radial minimum decreases from
$7.87\%$ to $1.99\%$ of the respective probe peak, compared
with $1.74\%$ for the diffraction-limited reference.

We model the circulating mode as an optical standing wave of an
$\mathrm{LG}_{0}^{1}$ mode
(see Fig.~\ref{fig:I donut and phase}(a) and
Eq.~\ref{eq:LG mode of the cavity}).
The corresponding laser parameters are the wavelength
$\lambda_\mathrm{L}=\quantity{1064}{nm}$, the one-way circulating
power $P=\quantity{34}{kW}$, and the focal waist
$w_\mathrm{f}=\quantity{8.1}{\text{\textmu}m}$.
The parameters used for this correction
example are summarized in
Table~\ref{tab:correction_parameters}.

\begin{table}[htbp]
\centering
\caption{Parameters used for the monochromatic
spherical-aberration-correction example with
$C_\mathrm{s}=\quantity{1}{mm}$ and a semi-convergence angle of
$\quantity{25}{mrad}$.
Here, $E_\mathrm{e}$ is the electron kinetic energy,
$\lambda_\mathrm{L}$ is the optical wavelength,
$w_\mathrm{f}$ is the optical waist radius,
$d_\mathrm{int}$ is the axial integration length used to
calculate the ponderomotive phase,
$M$ is the transverse magnification from the interaction
plane to the objective aperture plane,
$\max|\phi_\mathrm{p}|$ is the maximum absolute phase
over the full transverse profile shown in
Fig.~\ref{fig:I donut and phase}(b),
and $P$ is the required one-way circulating power.}
\label{tab:correction_parameters}
\small
\setlength{\tabcolsep}{8pt}
\begin{tabular}{ccccccc}
\toprule
$E_\mathrm{e}$ & $\lambda_\mathrm{L}$ & $w_\mathrm{f}$
& $d_\mathrm{int}$ & $M$ & $\max|\phi_\mathrm{p}|$ & $P$ \\
\midrule
$\quantity{5}{keV}$ &
$\quantity{1064}{nm}$ &
$\quantity{8.1}{\text{\textmu}m}$ &
$\approx\quantity{1.2}{mm}$ &
$\approx 4.2$ &
$\approx\quantity{39}{rad}$ &
$\quantity{34}{kW}$ \\
\bottomrule
\end{tabular}
\end{table}

Finite-energy-spread effects are evaluated by incoherent
wave-optical averaging over a Gaussian electron-energy
distribution, as described in Appendix~\ref{app:pupil}.
The ponderomotive spherical-aberration correction and
nominal defocus are held fixed, while chromatic aberration
is included with $C_c=\quantity{1}{mm}$.
For $E_e=\quantity{5}{keV}$,
$\theta_{\mathrm{max}}=\quantity{25}{mrad}$, and a representative
monochromated energy-distribution FWHM of
$\Delta E=\quantity{0.1}{eV}$~\cite{ADRIAANS2026114297},
the corrected probe has a diameter enclosing 50\% of the
total electron current (FW50) of \quantity{0.4572}{nm},
compared with \quantity{0.4303}{nm} in the monochromatic limit.
The FW50 therefore increases by approximately 6.3\%,
while the peak intensity decreases by approximately 6.5\%
relative to the $C_\mathrm{s}$-corrected monochromatic probe.
The peak remains at approximately 77.2\% of the
monochromatic diffraction-limited reference at equal
total electron probability.

Larger energy spreads increasingly redistribute electron
current towards larger radii and reduce the peak intensity.
At $\Delta E=\quantity{0.2}{eV}$, the FW50 is
\quantity{0.5424}{nm}, whereas at $\Delta E=\quantity{0.5}{eV}$
it reaches \quantity{1.059}{nm}, with a peak-intensity ratio
of approximately 0.397.
These results indicate modest chromatic degradation at
$\Delta E=\quantity{0.1}{eV}$ under the assumed operating
conditions, while demonstrating the importance of energy
spread for retaining a concentrated corrected probe.
The radial profiles, encircled-current fractions, and
quantitative metrics are presented in
Appendix~\ref{subsec:finite_energy_probe}.
Finite source size and higher-order lens aberrations
are neglected in these calculations.

The same framework can be extended to other aberrations
and target phase profiles by computing $\phi_\mathrm{p}$
for different intracavity standing-wave modes using the
analytical vector-field expressions of
Ref.~\cite{lembessis2024miniature} and optimizing their
coherent superposition subject to the cavity modal spectrum.

\section{Proposed experimental realization}\label{sec:level4}
The practical viability of the proposed cavity geometry is ultimately set by whether a near-concentric resonator with a central mirror aperture can still sustain the circulating power required for a sizeable ponderomotive phase shift. Although near-concentric operation provides the tight focal waist needed for strong electron–light coupling, the on-axis hole introduced for electron transmission can diffract and clip the intracavity mode, thereby reducing the resonant power enhancement and potentially becoming the main bottleneck of the concept. To quantify this effect, we go beyond analytic cavity estimates and perform FFT-based wave-propagation simulations of the intracavity field, including diffraction, aperture losses, and round-trip convergence to the steady state. These simulations are implemented using a modified version of the OSCAR package~\cite{degallaix2020oscar}; full simulation details are provided in appendix~\ref{app:cavity}.

\begin{figure}[t]
    \centering
    \includegraphics[width=10 cm]{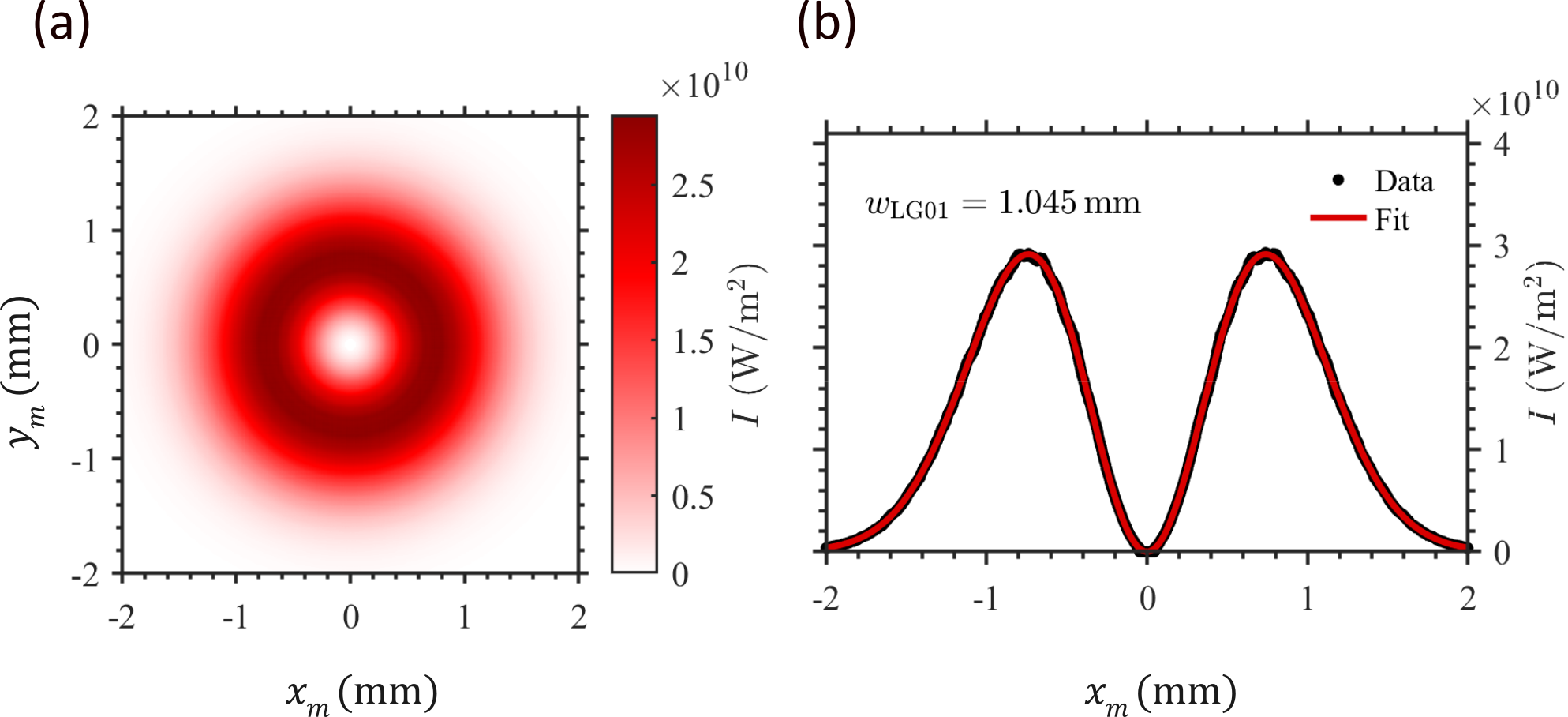}
    \caption{(a) Circulating intensity profile at approximately
$\quantity{130}{kW}$ at the inner side of the mirrors after
$10^{4}$ round trips. Both cavity mirrors have identical,
coaxial central holes of diameter
$D_\mathrm{h}=\quantity{100}{\text{\textmu}m}$. The transverse units are given by $x_\mathrm{m}$ and $y_\mathrm{m}$. (b) Horizontal cut of the circulating intensity at $y_m=0$, together with a fit to an $\mathrm{LG}_{0}^{1}$ intensity distribution. The fitted profile, with waist parameter $w_{\mathrm{fit}}=1.045~\mathrm{mm}$, closely reproduces the numerical result, indicating that the presence of the central hole only minimally perturbs the circulating $\mathrm{LG}_{0}^{1}$ mode.}
    \label{fig:Donut hole}
\end{figure}
\begin{figure}[t]
    \centering
    \includegraphics[width=6cm]{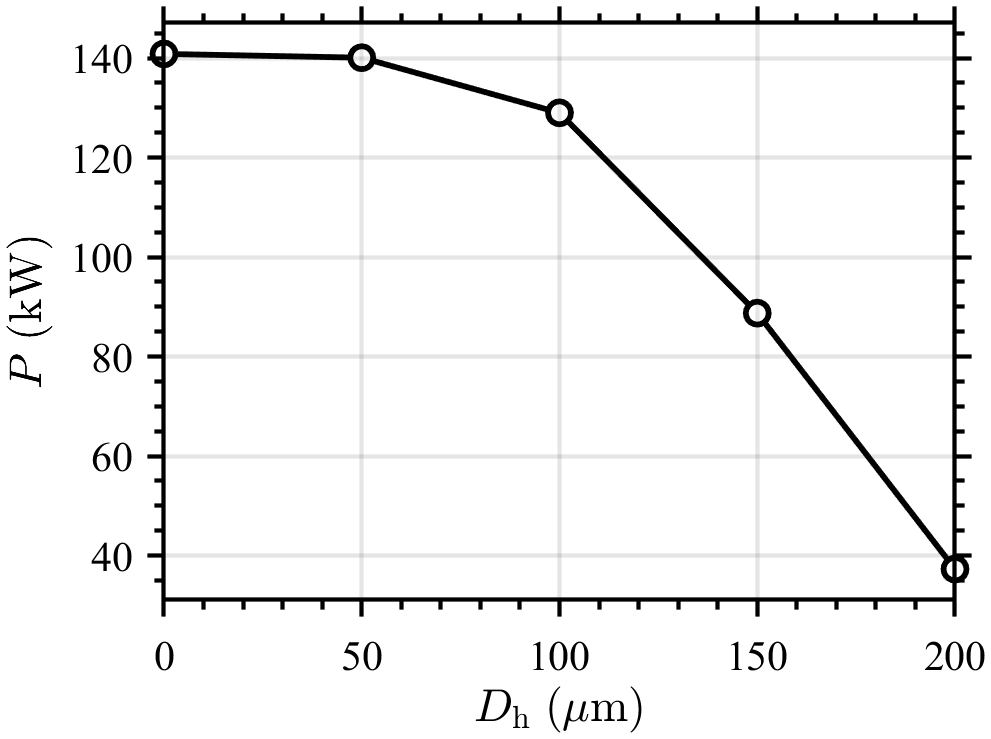}
    \caption{One-way circulating power $P$ in the Fabry--P\'erot cavity
as a function of the common central-hole diameter
$D_{\mathrm{h}}$, varied simultaneously in both mirrors. The two
holes are identical and coaxial. The power is evaluated after
$10^{4}$ round trips for an input power
$P_{\mathrm{in}}=\quantity{40}{W}$. The circulating power remains nearly
unchanged for small hole diameters, indicating that the
$\mathrm{LG}_{0}^{1}$ intracavity mode is only weakly perturbed.
For larger holes, however, the overlap of the mode with
the mirror apertures is progressively reduced,
resulting in a pronounced decrease in the circulating power.}
    \label{fig:Power vs hole diameter}
\end{figure}

We consider a modified near-concentric Fabry--P\'erot resonator that supports high circulating power in a tightly focused standing-wave mode. Both cavity mirrors contain identical, coaxial circular apertures
of diameter $D_\mathrm{h}=\quantity{100}{\text{\textmu}m}$, forming a clear
channel for electron transmission through the cavity
(see Fig.~\ref{fig:fig2}). In Fig.~\ref{fig:Power vs hole diameter},
this common aperture diameter is varied simultaneously for both mirrors. The cavity is formed by two identical high-curvature spherical or parabolic mirrors with radii of curvature $R_1=R_2=R=25~\mathrm{mm}$ and length $L=50~\mathrm{mm}-\delta z $, where $\delta z= \quantity{3}{\text{\textmu}m}$ sets the offset from the concentric point. In this symmetric geometry, the stability condition is $0<g^2<1$ with $g=1-L/R$. Operating at a small $\delta z$ yields a small waist near the center of the cavity, maximizing the strength of the ponderomotive interaction over an extended axial region. Operation under near-concentric conditions enables tight focusing and at the same time an increase of the beam size on the mirrors, thus lowering the peak intensity~\cite{kogelnik1966laser}:

\begin{equation}
w_\mathrm{m}^2=\frac{\lambda_L L}{\pi}\,\frac{1}{\sqrt{\left(\frac{L}{R}\right)\left(2-\frac{L}{R}\right)}},
\label{eq:beam waist at the mirror}
\end{equation}
For a symmetric near-concentric cavity, the beam waist at the cavity focus is given by~\cite{kogelnik1966laser}:
\begin{equation}
w_\mathrm{f}^2=\frac{\lambda_\mathrm{L}}{2\pi}\sqrt{L(2R-L)},
\label{eq:beam waist at focus}
\end{equation}
where $w_\mathrm{f}$ denotes the $1/e^2$ intensity radius at the waist for a Gaussian beam. As the cavity approaches the concentric limit $L\to 2R$, the focal waist decreases, while the beam radius on the mirrors increases according to Eq.~(\ref{eq:beam waist at the mirror}).

Figures~\ref{fig:Donut hole} and~\ref{fig:Power vs hole diameter} summarize the main results of the light wave-propagation analysis in the modified Fabry--P\'erot cavity. Figure~\ref{fig:Donut hole} examines the intracavity field for a representative central hole diameter of $D_\mathrm{h}=\quantity{100}{\text{\textmu}m}$ and shows that the circulating mode remains closely $\mathrm{LG}_{0}^{1}$-like despite the aperture. Figure~\ref{fig:Power vs hole diameter} quantifies the reduction in one-way circulating power with increasing hole diameter. Importantly, for the representative aperture $D_\mathrm{h}=\quantity{100}{\text{\textmu}m}$, the cavity still sustains a one-way circulating power of about~\quantity{130}{kW} ($\sim 3250$ power enhancement), well above the $\sim$\quantity{34}{kW} required for the spherical-aberration-correction example. This operating point assumes cavity-mirror reflectivities
of approximately $R_1^{(\mathrm{pow})}=0.99921$ for the in-coupling mirror
and $R_2^{(\mathrm{pow})}=0.99992$ for the second cavity mirror.  For the assumed mirror geometry and losses, the numerical
results indicate that the perforated cavity can sustain
circulating powers exceeding those required for the
spherical-aberration-correction example. The power enhancement could be increased by optimizing
the input-coupler transmission relative to the remaining
round-trip losses and reducing parasitic losses.
Absorption-induced thermoelastic deformation can modify mirror curvature and the 
cavity-mode waist, and can limit the attainable focal intensity in near-concentric 
resonators~\cite{axelrod2026thermoelastic}. Treating the full 40~ppm parasitic loss per mirror specified 
in Appendix~C.3 as absorption gives conservative absorbed-power estimates of 
approximately 1.36~W and 5.2~W per mirror at one-way circulating powers of 
34~kW and 130~kW, respectively. The optical simulations assume fixed mirror 
geometry and do not include the resulting thermoelastic deformation or its 
effect on the cavity mode; consequently, no quantitative claim of thermal 
feasibility is made. This motivates low-absorption coatings and characterization 
of the mode profile at the intended operating power.

A longer optical wavelength and operation closer to the concentric limit at a fixed cavity length would further relax the constraint on the diameter of the central hole.

A proposed arrangement for integration into an SEM column,
including vacuum compatibility, access to an aperture-conjugate
plane, and estimated alignment and stability requirements,
is discussed in Appendix~\ref{app:sem_integration}.

\section{Discussion and conclusion}\label{sec:level5}

A cavity-enhanced free-space phase plate for continuous electron beams is enabled by the ponderomotive interaction with a structured intracavity standing wave. In an on-axis near-concentric Fabry--P\'erot geometry, the electron beam experiences a cylindrically symmetric interaction over an extended distance, enabling large accumulated phase shifts at circulating powers accessible with established near-concentric resonator technology. The realizable phase profiles are set by the subset of optical modes supported by the cavity in the presence of the central hole. Within this constraint, we showed that the support of an intracavity LG$_0^1$-like mode, even in the presence of the central mirror aperture, enables generation of a compensating phase with a third-order spherical-aberration coefficient opposite in sign to that of a conventional electron lens, thereby fully compensating \(C_\mathrm{s}=\quantity{1}{mm}\) and yielding a near-diffraction-limited probe for a \quantity{5}{keV} monochromatic beam at a semi-convergence angle of \quantity{25}{mrad}.

A central advantage of the present approach is that the phase modulation is imparted in free space rather than in matter, thereby avoiding the electron loss, charging, contamination, and inelastic scattering that limit material phase plates. Moreover, in contrast to geometries in which the optical field is oriented perpendicular to the electron beam and the interaction length is set by the optical waist~\cite{schwartz2019laser}, the near-concentric cavity in our work provides an extended on-axis interaction region. Additionally, the proposed on-axis design remains viable at larger optical waists. By operating slightly away from the concentric point, the weaker field of a larger waist is offset by a longer interaction region, enabling comparable phase accumulation. The on-axis geometry also avoids the need for multiple sequential cavities envisioned in crossed-cavity schemes~\cite{uesugi2025crossed} and may provide a less complex alternative to conventional multipole aberration correctors~\cite{haider2009current}.

The method is nevertheless constrained by three main considerations. First, the attainable transverse phase profiles are limited by the set of cavity eigenmodes that can be supported. Consequently, arbitrary phase patterns demonstrated in pulsed free-space beam-shaping geometries~\cite{chirita2022transverse} do not directly translate to the continuous-wave cavity setting, and practical wavefront synthesis must be formulated as an optimization within a discrete modal basis. Second, practical probe performance depends
on the electron-energy spread and the uncorrected chromatic
aberration, as quantified in Appendix~\ref{app:pupil}.

Third, the high cavity $Q$ factor ($Q\sim6.5\times10^8$),
corresponding to a photon lifetime of approximately
\quantity{370}{ns}, provides temporal filtering of
rapid laser-power fluctuations. Slower power variations
and changes in incident-beam pointing can still affect
the circulating power through changes in input power
and mode coupling. Stable phase modulation therefore
requires sufficient laser-power stability, stable mode
matching, and control of cavity detuning and alignment
relative to the electron beam, as discussed in
Appendix~\ref{app:sem_integration}. 

The high circulating power required here motivates careful consideration of stability and noise, including Johnson-noise-induced phase fluctuations~\cite{uhlemann2013thermal}. Longer-wavelength single-frequency fiber lasers would further lower the intensity required for a given phase shift and could be traded for larger electron semi-angles or reduced circulating power. More generally, for the present application, using a first-order Laguerre--Gaussian mode reduces the peak intensity at the cavity mirrors relative to a Gaussian mode at the same circulating power, thereby mitigating local heating and easing thermal-stability constraints. In addition, the required circulating power is only $\sim$\quantity{34}{kW}, a factor of four lower than that reported in \cite{turnbaugh2021high}. Given that multi-megawatt circulating powers have recently been demonstrated \cite{arakcheev2026probing}, the modulation scheme described in the present manuscript could potentially be extended to electron energies typical of TEMs.

A possible extension is cavity-based phase contrast using a donut-shaped optical mode to shift the scattered electron wave relative to the unscattered reference. Higher-order Laguerre–Gaussian modes are candidates, but their nonuniform radial intensity requires numerical assessment of the induced ponderomotive phase and resulting contrast transfer.

This reconfigurable cavity platform
provides a route to optical $C_s$ correction of continuous
electron beams at SEM-relevant energies and experimentally
accessible power levels.
Near-diffraction-limited focal probes are predicted in
the monochromatic limit.
For a Gaussian energy-distribution FWHM of
$\Delta E=\quantity{0.1}{eV}$ and $C_c=\quantity{1}{mm}$,
incoherent wave-optical averaging predicts a diameter
enclosing 50\% of the electron current of approximately
\quantity{0.46}{nm} at \quantity{5}{keV} and \quantity{25}{mrad}.
Larger energy spreads reduce the peak intensity and
current concentration, motivating monochromation
when targeting the smallest corrected probes. The electron-modulation cavity concept could be validated in an SEM equipped with a high-spatial-resolution electron detector~\cite{ixquiac2025low}.

\section*{Acknowledgements}
The authors thank NSL Stre{\v{s}}kov{\'a} for fruitful discussions.

\section*{Funding}
M.C.C.M. acknowledges support from the project MSCA Fellowships CZ--UK4 (CZ.02.01.01/\allowbreak 00/\allowbreak 22\_010/\allowbreak 0013392) under the Programme Johannes Amos Comenius. M.K. acknowledges support from the Czech Science Foundation (project 22-13001K), the European Union (ERC, eWaveShaper, 101039339), and the Ministry of Education, Youth and Sports of the Czech Republic (TERAFIT project No.\ CZ.02.01.01/\allowbreak 00/\allowbreak 22\_008/\allowbreak 0004594, funded by OP JAK, call Excellent Research).

\section*{Author contributions}
MCCM and MK developed the concept and contributed to the interpretation of the results. MCCM performed the analytical calculations and numerical simulations. MCCM led the cavity simulations, with contributions from JS. MCCM wrote the initial manuscript draft. MCCM and MK reviewed and revised the manuscript.

\section*{Data availability}
The data supporting the findings of this study are openly available at \cite{ChiritaMihaila2026ZenodoCavity}.

\section*{Competing interests}
The method for coaxial optical-cavity phase modulation of continuous electron beams described in this manuscript is the subject of international patent application No.\ PCT/CZ2026/050035. The patent applicant is Univerzita Karlova / Charles University. The inventors are M.C.C.M. and M.K. The international filing date was 8 June 2026.

\FloatBarrier

\newcounter{appsec}
\renewcommand{\theappsec}{\Alph{appsec}}
\newcounter{appsubsec}[appsec]
\renewcommand{\theappsubsec}{\theappsec.\arabic{appsubsec}}

\renewcommand{\theHequation}{appendix.\arabic{appsec}.\arabic{equation}}
\renewcommand{\theHfigure}{appendix.\arabic{appsec}.\arabic{figure}}
\renewcommand{\theHtable}{appendix.\arabic{appsec}.\arabic{table}}
\newcommand{\appendixsection}[2]{%
  \refstepcounter{appsec}%
  \setcounter{appsubsec}{0}%
  \setcounter{equation}{0}%
  \setcounter{figure}{0}%
  \setcounter{table}{0}%
  \renewcommand{\theequation}{\theappsec\arabic{equation}}%
  \renewcommand{\thefigure}{\theappsec\arabic{figure}}%
  \renewcommand{\thetable}{\theappsec\arabic{table}}%
  \section*{Appendix \theappsec. #1}%
  \label{#2}%
}
\newcommand{\appendixsubsection}[2]{%
  \refstepcounter{appsubsec}%
  \subsection*{\theappsubsec\quad #1}%
  \label{#2}%
}

\appendixsection{Electric field of the standing wave}{app:field}

Within the cavity, the two counter-propagating beams interfere to form a standing wave, with the total electric field is described by ~\cite{lembessis2024miniature}:
\begin{equation}
\begin{aligned}
\bm{E}(\rho, \varphi, z, t) =\ 
&\mathcal{E}_0\, e^{i(\ell \varphi - \omega_0 t)} \cos(k_\mathrm{L}z + \Theta)\, u(\rho, z) \\
&\quad\times\left[
(\alpha\, \hat{x} + \beta\, \hat{y}) 
+ i \cdot \frac{\rho z}{z^2 + z_R^2} 
(\alpha \cos\varphi + \beta \sin\varphi)
\hat{z}
\right] \\
&+ \mathcal{E}_0\, e^{i(\ell \varphi - \omega_0 t)} \cdot \frac{\sin(k_\mathrm{L}z + \Theta)}{k_\mathrm{L}} \\
&\quad\times\left[
- (\alpha \cos\varphi + \beta \sin\varphi) \frac{\partial u}{\partial \rho}
- i \cdot \frac{\ell u}{\rho} (\beta \cos\varphi - \alpha \sin\varphi)
\right]
\hat{z},
\end{aligned}
\label{eq:LG mode of the cavity}
\end{equation}
where the phase \(\Theta(\rho, z)\) is given by

\begin{equation}
\Theta(\rho, z) = -(2p + |\ell| + 1) \arctan\left(\frac{z}{z_R}\right) + \frac{k_\mathrm{L} \rho^2 z}{2 (z^2 + z_R^2)} \,,
\end{equation}

where \(\rho\), \(\varphi\), and \(z\) are cylindrical coordinates with \(z\) along the cavity axis, \(\hat{x}\), \(\hat{y}\), and \(\hat{z}\) are the Cartesian unit vectors, \(\mathcal{E}_0\) is the field amplitude, \(\omega_0\) is the angular frequency, \(k_\mathrm{L}\) is the wave number. The complex coefficients \(\alpha\) and \(\beta\) define the common transverse polarization state of the two counter-propagating beams. In the present case, the field is linearly polarized along the \(x\)-direction, corresponding to \(\alpha = 1\) and \(\beta = 0\). The same azimuthal index \(\ell\) is assigned to both beams, \(\ell_+ = \ell_- = \ell\), so \(\ell\) characterizes the orbital angular momentum winding number of each Laguerre--Gaussian mode. The transverse mode structure is captured by \(u(\rho, z)\), given by:

\begin{equation}
u(\rho, z) = \frac{C_{|\ell| p}}{\sqrt{1 + z^2 / z_R^2}} 
\left( \frac{\rho \sqrt{2}}{w_\mathrm{f} \sqrt{1 + z^2 / z_R^2}} \right)^{|\ell|} 
\exp\left[ -\frac{\rho^2}{w_\mathrm{f}^2 (1 + z^2 / z_R^2)} \right] 
L_p^{(|\ell|)}\left( \frac{2 \rho^2}{w_\mathrm{f}^2 (1 + z^2 / z_R^2)} \right),
\end{equation}
where \(w_\mathrm{f}\) is the beam waist in focus, \(z_R = \pi w_\mathrm{f}^2 / \lambda_\mathrm{L}\) is the Rayleigh range, \(L_p^{(|\ell|)}\) is the generalized Laguerre polynomial of the radial index \(p\), and $C_{|\ell|,p} = \sqrt{p!/(p + |\ell|)!}$. This formulation fully describes the spatial and polarization structure of the standing-wave field inside the cavity, including both transverse and longitudinal components.

\appendixsection{Pupil-plane phases and finite-energy-spread probe formation}{app:pupil}

\appendixsubsection{Pupil-plane lens aberration phases}{subsec:pupil_phases}

\begin{figure}[t]
    \centering
    \includegraphics[width=\textwidth]{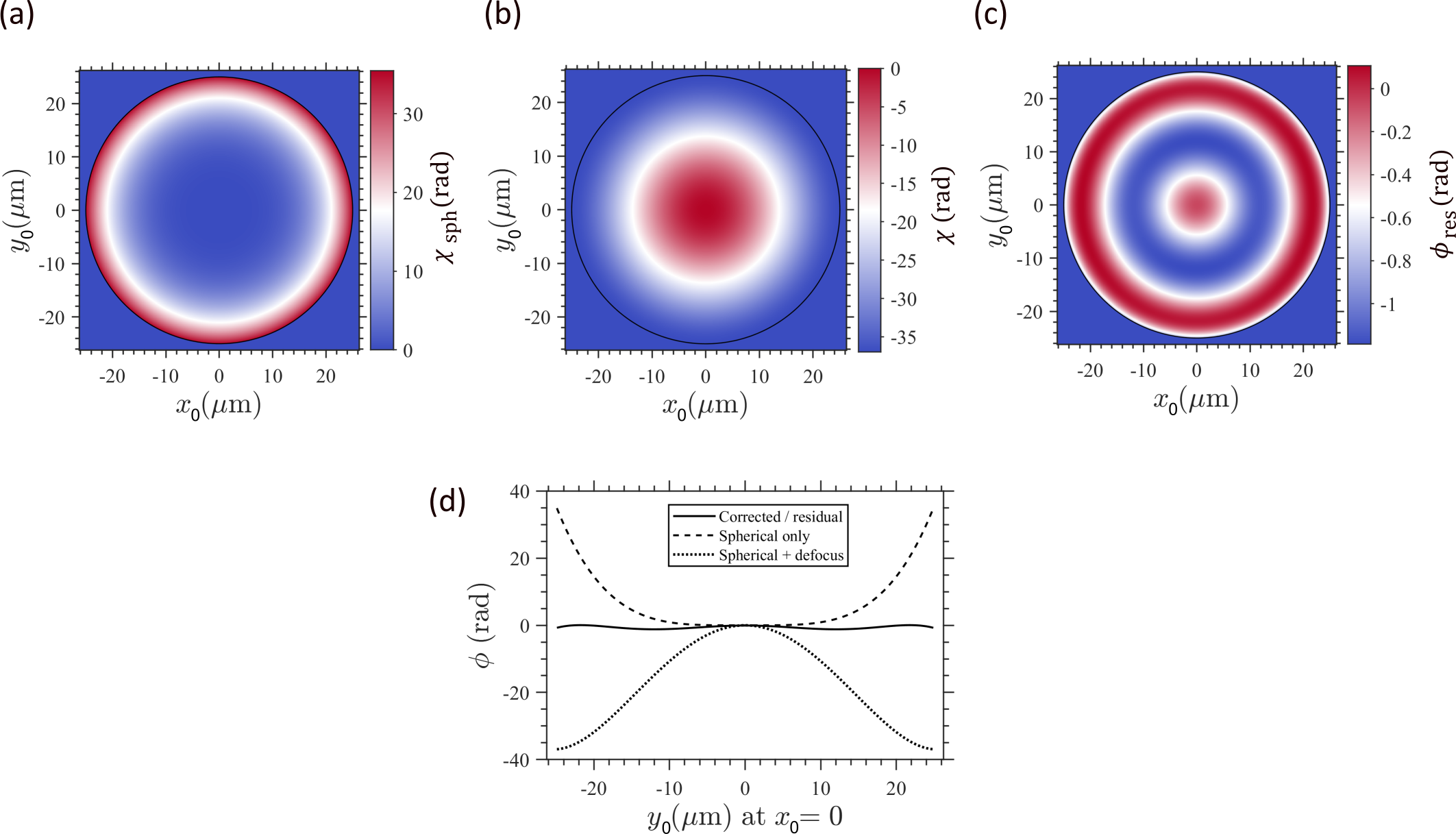}
    \caption{Pupil-plane phase maps used to characterize the lens
    aberrations. (a) Spherical-aberration phase
    $\chi_{\mathrm{sph}}$, (b) total lens aberration phase $\chi$
    including spherical aberration and defocus, and (c) residual
    phase $\phi_{\mathrm{res}}$ after ponderomotive correction.
    (d) Corresponding line profiles extracted along $x_0=0$,
    comparing the corrected residual phase, spherical-only
    contribution, and total spherical-plus-defocus phase.}
    \label{fig:pupil-phase-maps}
\end{figure}

To clarify the phase contributions entering the electron probe formation,
we explicitly decompose the lens aberration phase in the pupil plane.
The total aberration phase used in the main text is written as
\begin{equation}
\chi(\theta)
=
\frac{\pi}{2\lambda_e}
\left(C_s\theta^4-2\Delta z\theta^2\right).
\label{eq:app_lens_phase}
\end{equation}
The spherical-aberration-only contribution is therefore
\begin{equation}
\chi_{\mathrm{sph}}(\theta)
=
\frac{\pi}{2\lambda_e}C_s\theta^4.
\label{eq:app_spherical_phase}
\end{equation}
The residual phase after ponderomotive correction is defined as
\begin{equation}
\phi_{\mathrm{res}}
=
\phi_{\mathrm{p},\mathrm{obj}}-\chi,
\label{eq:app_residual_phase}
\end{equation}
where $\phi_{\mathrm{p},\mathrm{obj}}$ denotes the ponderomotive phase
imparted by the optical field, expressed in the coordinate system
of the objective-lens pupil plane.
Figure~\ref{fig:pupil-phase-maps} shows the individual phase maps
and the corresponding line profiles through the pupil center,
illustrating how the ponderomotive phase compensates the dominant
lens-induced aberration.

\appendixsubsection{Finite-energy-spread probe formation}{subsec:finite_energy_probe}

To evaluate the effect of chromatic aberration  of the objective lens, we extend the
monochromatic probe calculation by incoherently averaging the
focal-plane intensities over the electron-energy
distribution~\cite{chirita2025design}.
We assume a normalized Gaussian distribution of energy deviations
$\delta E$ from the mean kinetic energy $E_e$:
\begin{equation}
g(\delta E)
=
\frac{1}{\sqrt{2\pi}\sigma_E}
\exp\!\left(-\frac{\delta E^2}{2\sigma_E^2}\right),
\qquad
\sigma_E=\frac{\Delta E}{2\sqrt{2\ln 2}},
\label{eq:app_energy_distribution}
\end{equation}
where $\Delta E$ is the energy-distribution FWHM.
The monochromatic limit $\Delta E=0$ is evaluated separately.

Within the first-order chromatic-defocus approximation,
the defocus entering Eq.~\eqref{eq:app_lens_phase} becomes
\begin{equation}
\Delta z(\delta E)
=
\Delta z_0+C_c\frac{\delta E}{E_e},
\label{eq:app_chromatic_defocus}
\end{equation}
where $C_c$ is the chromatic-aberration coefficient and
$\Delta z_0$ is the nominal defocus at the mean energy.
The pupil, electron wavelength, and correcting ponderomotive
phase are evaluated at the mean energy and held fixed;
the energy spread is included through its chromatic defocus.
For each energy deviation, the probe wave function
$\psi_f(x_f,y_f;\delta E)$ is calculated using
Eqs.~\eqref{eq:pupil} and \eqref{eq:probe_fft}.
The energy-averaged intensity is then
\begin{equation}
\overline{I}(x_f,y_f)
=
\int_{-\infty}^{\infty}
g(\delta E)
\left|\psi_f(x_f,y_f;\delta E)\right|^2
\,\mathrm{d}\delta E.
\label{eq:app_energy_averaged_intensity}
\end{equation}

The calculations use $E_e=\quantity{5}{keV}$,
$\theta_{\mathrm{max}}=\quantity{25}{mrad}$,
$C_s=C_c=\quantity{1}{mm}$, and
$\Delta z_0=\quantity{638}{nm}$, with the same optical correction
as in the monochromatic example.
Finite source size and higher-order lens aberrations are neglected.
All probes are normalized to equal total electron probability.

To characterize the spatial concentration of electron current,
we calculate the encircled-current fraction
\begin{equation}
F_{\mathrm{enc}}(R)
=
\frac{
\displaystyle
\iint_{x_f^2+y_f^2\leq R^2}
\overline{I}(x_f,y_f)\,
\mathrm{d}x_f\,\mathrm{d}y_f
}{
\displaystyle
\iint
\overline{I}(x_f,y_f)\,
\mathrm{d}x_f\,\mathrm{d}y_f
}.
\label{eq:app_encircled_current}
\end{equation}
The diameter enclosing 50\% of the total electron current is
defined as $\mathrm{FW50}=2R_{50}$, where
$F_{\mathrm{enc}}(R_{50})=0.5$.
The peak-intensity ratio is
$\overline{I}_{\mathrm{max}}/I_{\mathrm{DL,max}}$,
where $I_{\mathrm{DL,max}}$ is the peak intensity of the
monochromatic diffraction-limited reference for the same pupil
and total electron probability.

\begin{figure}[t]
    \centering
    \includegraphics[width=12cm]
    {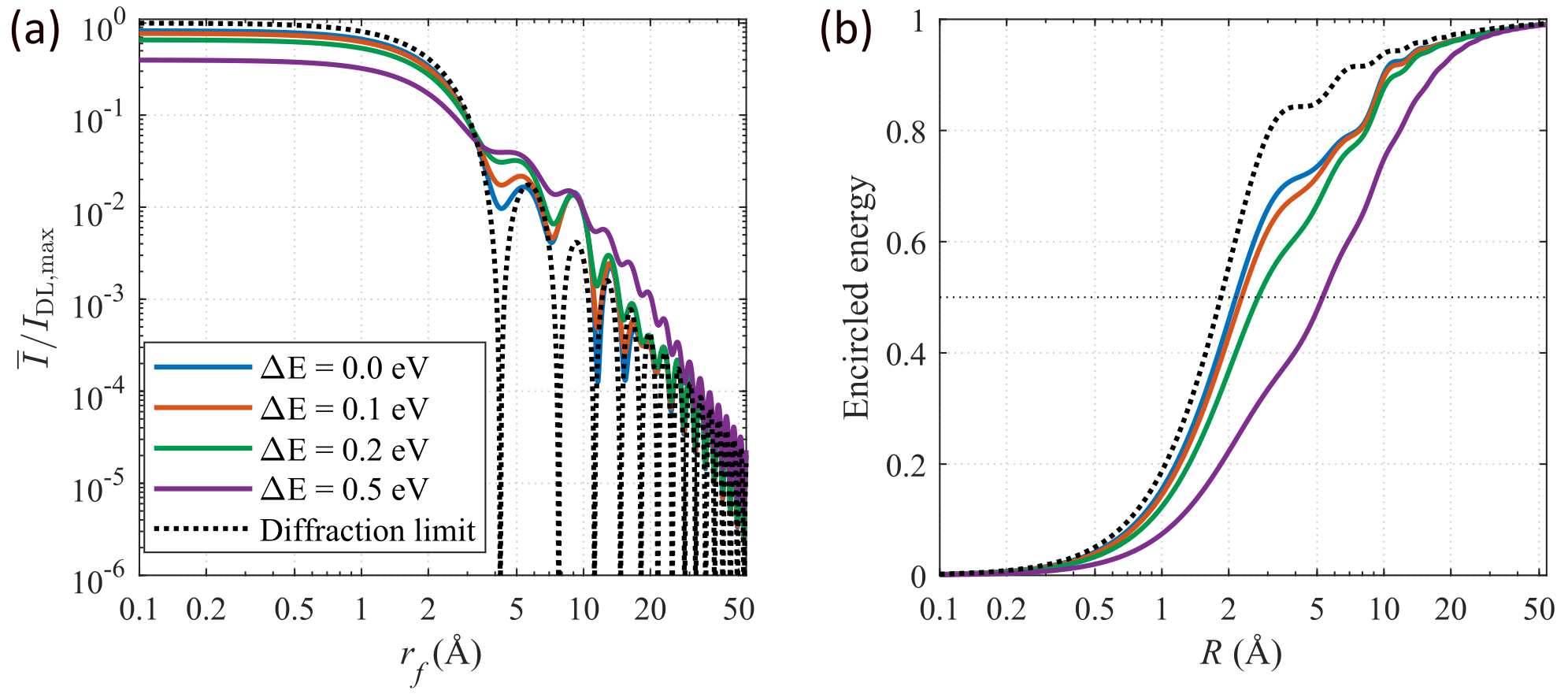}
    \caption{Finite-energy-spread performance of the
    ponderomotively corrected electron probe for
    $E_e=\quantity{5}{keV}$,
    $\theta_{\mathrm{max}}=\quantity{25}{mrad}$,
    $C_s=C_c=\quantity{1}{mm}$, and
    nominal defocus $\Delta z_0=\quantity{638}{nm}$.
    Left: azimuthally averaged focal-plane intensity,
    normalized to the peak of the monochromatic
    diffraction-limited reference and displayed on
    logarithmic axes.
    Right: encircled-current fraction as a function of radius.
    Solid curves correspond to Gaussian energy distributions
    with FWHM $\Delta E=0$, 0.1, 0.2, and \quantity{0.5}{eV},
    using the same fixed optical correction and nominal defocus.
    The dotted black curve denotes the monochromatic
    diffraction-limited reference.
    All probes use the same uniform circular pupil and are
    normalized to equal total electron probability.
    The horizontal dotted line in the right panel marks
    the 50\% current fraction used to determine the FW50.}
    \label{fig:chromatic_probe_profiles}
\end{figure}

Figure~\ref{fig:chromatic_probe_profiles} compares the radial
intensity distributions and encircled-current fractions for
energy spreads from 0 to \quantity{0.5}{eV}.
The corresponding quantitative results are summarized in
Table~\ref{tab:chromatic_probe_metrics}.
At $\Delta E=\quantity{0.1}{eV}$, the corrected probe has
$\mathrm{FW50}=\quantity{0.4572}{nm}$ and a peak-intensity ratio
of 0.7724.
Relative to the corrected monochromatic probe, the FW50
increases by approximately 6.3\%, while the peak intensity
decreases by approximately 6.5\%.

Increasing the energy spread redistributes electron current
towards larger radii and reduces the peak intensity.
At $\Delta E=\quantity{0.5}{eV}$, the FW50 reaches
\quantity{1.059}{nm}, and the peak-intensity ratio decreases to 0.3968.
The wave-optical calculation therefore quantifies both the
retained current concentration at small energy spread and
the increasing chromatic degradation at larger spreads.

\begin{table}[t]
    \centering
    \caption{Wave-optical probe metrics for the operating
    point in Fig.~\ref{fig:chromatic_probe_profiles}.
    For probes with ponderomotive $C_s$ correction, the chromatic-aberration coefficient is fixed at $C_c=\quantity{1}{mm}$.
    $\Delta E$ denotes the energy-distribution FWHM.
    The peak-intensity ratio is referenced to the
    monochromatic diffraction-limited probe at equal
    total electron probability.
    FW50 is the diameter enclosing 50\% of the total
    electron current.}
    \label{tab:chromatic_probe_metrics}
    \begin{tabular}{lccc}
        \toprule
        Probe
        & $\Delta E$ (eV)
        & $\overline{I}_{\mathrm{max}}/I_{\mathrm{DL,max}}$
        & FW50 (nm) \\
        \midrule
        Diffraction limit & 0.0 & 1.0000 & 0.3685 \\
        $C_s$-Corrected         & 0.0 & 0.8261 & 0.4303 \\
        $C_s$-Corrected         & 0.1 & 0.7724 & 0.4572 \\
        $C_s$-Corrected         & 0.2 & 0.6573 & 0.5424 \\
        $C_s$-Corrected         & 0.5 & 0.3968 & 1.059  \\
        \bottomrule
    \end{tabular}
\end{table}

\appendixsection{Near-concentric Fabry--Perot cavity simulations}{app:cavity}

We simulate the modified near-concentric Fabry--Perot cavity introduced in the
main text to determine the intracavity mode shape and circulating
power in the presence of the on-axis mirror aperture used for electron
transmission.

\appendixsubsection{Cavity geometry and input field}{subsec:cavity_geometry}

We consider a symmetric near-concentric Fabry--Perot resonator formed by two
identical mirrors with radii of curvature
\begin{equation}
R_1 = R_2 = R = \quantity{25}{mm},
\end{equation}
and cavity length
\begin{equation}
L = \quantity{50}{mm} - \delta z,
\qquad
\delta z = \quantity{3}{\text{\textmu}m},
\end{equation}
so that the cavity operates close to the concentric limit. Each mirror has a
clear diameter
\begin{equation}
D_{\mathrm{mir}} = \quantity{6.8}{mm},
\end{equation}
and contains an identical, coaxial central circular aperture of diameter
\begin{equation}
D_\mathrm{h} = \quantity{100}{\text{\textmu}m}
\end{equation}
for on-axis electron transmission, unless otherwise stated.

For this symmetric geometry, the beam radius on the mirrors and the waist
radius at the cavity focus are~\cite{kogelnik1966laser}:
\begin{equation}
w_\mathrm{m}^2=
\frac{\lambda_\mathrm{L} L}{\pi \sqrt{(L/R)\left(2-L/R\right)}},
\label{eq:wm_supp}
\end{equation}
and
\begin{equation}
w_\mathrm{f}^2=
\frac{\lambda_\mathrm{L}}{2\pi}\sqrt{L(2R-L)},
\label{eq:wL_supp}
\end{equation}
respectively. For the parameters used here, these expressions give
\begin{equation}
w_\mathrm{m} \approx \quantity{1.045}{mm},
\qquad
w_\mathrm{f} \approx \quantity{8.10}{\text{\textmu}m}.
\end{equation}

Throughout this appendix, the optical fields are represented
by power-normalized complex amplitudes, with units
$\mathrm{W}^{1/2}\,\mathrm{m}^{-1}$. Their squared modulus
therefore gives optical intensity, and its transverse
integral gives the one-way optical power. The injected field is treated as a scalar \(LG_0^1\) spatial mode with a fixed transverse polarization,
\begin{equation}
\mathbf{E}_{\mathrm{in}}(\rho,\varphi)
=
\hat{\mathbf{e}}_{\mathrm{pol}}
E_0
\left(\frac{\sqrt{2}\rho}{w_{\mathrm{m}}}\right)
\exp\!\left(-\frac{\rho^2}{w_{\mathrm{m}}^2}\right)
e^{i\varphi},
\label{eq:LG01_input}
\end{equation}
where \(\hat{\mathbf{e}}_{\mathrm{pol}}\) is a constant transverse polarization unit vector and $E_0$ is the input-field amplitude, determined by the power normalization in Eq. (C9). The polarization state is assumed to be uniform across the beam and therefore does not affect the scalar cavity-coupling calculation. Here \((\rho,\varphi)\) are the transverse cylindrical coordinates, with \(\rho=\sqrt{x^2+y^2}\) and \(\varphi\) the azimuthal angle about the optical axis. The normalization is chosen such that
\begin{equation}
\iint |E_{\mathrm{in}}(x,y)|^2\,dx\,dy
=
P_{\mathrm{in}},
\end{equation}
where \(P_{\mathrm{in}}\) is the input laser power. The input coupling is described in section~\ref{subsec:mirror_operators}.

\appendixsubsection{FFT-based wave-propagation model}{subsec:fft_model}

The cavity field is computed with a modified version of OSCAR~\cite{degallaix2020oscar}, which
represents the optical field on a two-dimensional transverse grid and
propagates it between optical elements using Fresnel propagation based on FFT.
For propagation over a distance \(d\), the spatial-frequency representation of the field is multiplied by
\begin{equation}
\exp\!\left[
-i k_\mathrm{L} d + i\pi \lambda_\mathrm{L} (f_x^2+f_y^2)d
\right],
\label{eq:fresnel_operator}
\end{equation}
where \(f_x\) and \(f_y\) are the transverse spatial-frequency variables. The propagated field in real space is obtained by applying the inverse Fourier transform.

All optical elements are applied as multiplicative maps in real space. The
finite mirror aperture and the central electron hole are implemented through
the binary mask
\begin{equation}
M(\rho)=
\begin{cases}
1, & r_h \le \rho \le D_{\mathrm{mir}}/2,\\[4pt]
0, & \text{otherwise},
\end{cases}
\label{eq:mask_def}
\end{equation}
with
\begin{equation}
r_\mathrm{h}=\frac{D_\mathrm{h}}{2}.
\end{equation}

\appendixsubsection{Mirror operators and input coupling}{subsec:mirror_operators}

Inside the cavity, each mirror is modeled by its amplitude reflectivity, a
finite-aperture mask, and a curvature-induced phase shift. For a spherical mirror of radius of curvature $R_j$, where $j=\mathrm{ITM}$ denotes the input coupling mirror and $j=\mathrm{ETM}$ denotes the end mirror, the exact surface sag relative to the mirror vertex is
\begin{equation}
s_j(\rho)=R_j-\sqrt{R_j^2-\rho^2},
\end{equation}
which produces an optical path difference upon reflection
\begin{equation}
\Delta_j^{\mathrm{mir}}(\rho)=2s_j(\rho).
\end{equation}
In the paraxial limit $\rho \ll |R_j|$, this reduces to
\begin{equation}
\Delta_j^{\mathrm{mir}}(\rho)\simeq \frac{\rho^2}{R_j}.
\label{eq:paraxial_mirror_phase}
\end{equation}
The corresponding reflection operator is
\begin{equation}
\mathcal{R}_j(\rho)=
r_j\,M(\rho)\,
\exp\!\left[ik_\mathrm{L}\Delta_j^{\mathrm{mir}}(\rho)\right],
\label{eq:mirror_operator}
\end{equation}
where $r_j$ is the mirror amplitude reflectivity. The same mask $M(\rho)$ is applied at every
reflection from both the input coupling mirror (ITM) and the
end mirror (ETM), so each round trip includes one encounter
with each aperture. Although the physical
resonator employs spherical mirrors, the intracavity field evolution is
computed within a paraxial Fresnel/FFT framework. To preserve consistency
within this approximation, the mirror reflection phase is implemented using
the quadratic form (see Eq.~\eqref{eq:paraxial_mirror_phase}), rather than the
exact spherical expression. In the near-concentric regime, particularly for
small $\delta z<\quantity{40}{\text{\textmu}m}$, the use of the exact spherical phase within the paraxial numerical
scheme introduces higher-order nonparaxial contributions that are not
consistently resolved on the numerical grid and lead to poor numerical
convergence. A fully nonparaxial treatment would require field
propagation to be described within a vectorial diffraction framework~\cite{richards1959electromagnetic,dorn2003sharper}, in which the exact spherical mirror phase could be treated consistently, albeit at increased computational cost.

To improve mode matching in the near-concentric regime, the input cavity mirror is designed as a combined refractive--reflective element. Its entrance surface is shaped to provide an anaclastic phase precompensation, while the inner, cavity-facing surface carries the high-reflectivity coating. The injected field is therefore precompensated by the anaclastic entrance phase profile~\cite{durak2014diffraction}:
\begin{equation}
\phi_{\mathrm{ana}}(\rho)=
-k_\mathrm{L}(n-1)a
\left(
1-\sqrt{1+\frac{\rho^2}{b^2}}
\right),
\label{eq:anaclastic_phase}
\end{equation}
with
\begin{equation}
a=\frac{f_\mathrm{a}n}{n+1},
\qquad
b=f_\mathrm{a}\sqrt{\frac{n-1}{n+1}},
\qquad
f_\mathrm{a}=R+t_g.
\end{equation}
Here $n$ is the substrate refractive index, $t_g$ is the substrate thickness, and $f_\mathrm{a}$ is the focal length associated with the anaclastic entrance surface. The transmitted field at the input mirror is thus multiplied by the phase factor $\exp[i\phi_{\mathrm{ana}}(\rho)]$ before intracavity propagation and reflection from the cavity-facing mirror surface.

The mirror power coefficients used in the simulations are
\begin{equation}
T_{\mathrm{ITM}}=\quantity{750}{ppm},
\qquad
T_{\mathrm{ETM}}=\quantity{40}{ppm},
\end{equation}
with parasitic losses
\begin{equation}
L_{\mathrm{ITM}}=L_{\mathrm{ETM}}=\quantity{40}{ppm}.
\end{equation}
The corresponding power reflectivities are
\begin{equation}
R_j^{(\mathrm{pow})}=1-T_j-L_j,
\label{eq:mirror_power_reflectivity}
\end{equation}
and the amplitude coefficients are
\begin{equation}
r_j=\sqrt{R_j^{(\mathrm{pow})}},
\qquad
t_j=\sqrt{T_j}.
\label{eq:mirror_amplitude_coefficients}
\end{equation}

\appendixsubsection{Simulation parameters used for Figs.~5 and 6}{subsec:simulation_parameters}

Table~\ref{tab:cavity_params} summarizes the physical and numerical parameters
used for the cavity simulations shown in Figs.~5 and 6 of the main text. For
Fig.~6, only the central hole diameter $D_\mathrm{h}$ is varied, while all other
parameters are kept fixed. In our simulations we assumed $f_\mathrm{a}=\quantity{36.2}{mm}$ and $n=1.45$; however, one could alternatively consider $n=1.48$, corresponding to premium-grade ULE Corning 7972 glass, which offers a lower thermal expansion coefficient.

\begin{table}[t]
\centering
\caption{Physical and numerical parameters used in the cavity simulations.}
\label{tab:cavity_params}
\small
\begin{tabular}{lll}
\toprule
Quantity & Symbol & Value \\
\midrule
Laser wavelength & $\lambda_\mathrm{L}$ & \quantity{1064}{nm} \\
Input power & $P_{\mathrm{in}}$ & \quantity{40}{W} \\
Mirror radii of curvature & $R_1,R_2$ & \quantity{25}{mm}, \quantity{25}{mm} \\
Focal length of the anaclastic lens & $f_\mathrm{a}$ & \quantity{36.2}{mm} \\
Cavity length & $L$ & $\quantity{50}{mm}-\quantity{3}{\text{\textmu}m}$ \\
Offset from concentric point & $\delta z$ & \quantity{3}{\text{\textmu}m} \\
Mirror clear diameter & $D_{\mathrm{mir}}$ & \quantity{6.8}{mm} \\
Central hole radius & $r_\mathrm{h}$ & \quantity{50}{\text{\textmu}m} \\
Representative hole diameter & $D_\mathrm{h}$ & \quantity{100}{\text{\textmu}m} \\
ITM transmission & $T_{\mathrm{ITM}}$ & \quantity{750}{ppm} \\
ETM transmission & $T_{\mathrm{ETM}}$ & \quantity{40}{ppm} \\
ITM loss & $L_{\mathrm{ITM}}$ & \quantity{40}{ppm} \\
ETM loss & $L_{\mathrm{ETM}}$ & \quantity{40}{ppm} \\
Substrate refractive index & $n$ & 1.45 \\
Substrate thickness & $t_\mathrm{g}$ & \quantity{11.2}{mm} \\
Grid size & $N_x \times N_y$ & $1750 \times 1750$ \\
Sampling pitch & $\Delta x=\Delta y$ & \quantity{4}{\text{\textmu}m} \\
Transverse window size & $L_x \times L_y$ & \quantity{7.0}{mm}$\times$\quantity{7.0}{mm} \\
Injected mode & $LG_0^1$ & -- \\
Mirror-plane beam radius & $w_\mathrm{m}$ & \quantity{1.045}{mm} \\
Waist radius at focus & $w_\mathrm{f}$ & \quantity{8.10}{\text{\textmu}m} \\
\bottomrule
\end{tabular}
\end{table}

\appendixsubsection{Steady-state reconstruction and post-processing}{subsec:convergence_postprocessing}

The steady-state intracavity field is reconstructed by iterating the cavity
round-trip map and coherently summing the resulting field sequence. Denoting by
$\mathcal{P}(L)$ the Fresnel propagation operator over one cavity length, we
write
\begin{equation}
E_{q+1}=\mathcal{U}_{\mathrm{rt}}\,E_q,
\qquad
\mathcal{U}_{\mathrm{rt}}
\equiv
e^{i\phi_{\mathrm{rt}}}
\mathcal{R}_{\mathrm{ITM}}\,
\mathcal{P}(L)\,
\mathcal{R}_{\mathrm{ETM}}\,
\mathcal{P}(L),
\qquad
\phi_{\mathrm{rt}}=2k_L\Delta L_{\mathrm{res}},
\label{eq:roundtrip_update}
\end{equation}
where $q$ is the round-trip index, $E_q$ is the field contribution after $q$ round trips, $\mathcal{U}_{\mathrm{rt}}$ is the round-trip propagation operator, $\phi_{\mathrm{rt}}$ is the round-trip phase offset, and $\Delta L_{\mathrm{res}}$ is the microscopic cavity-length offset used to place the cavity on resonance. For the simulations shown in Figs.~5 and~6, we
used $\Delta L_{\mathrm{res}}=\quantity{0.709248}{\text{\textmu}m}$, obtained using
\cite{degallaix2020oscar}.

The circulating field reconstructed after $N$ round trips is
\begin{equation}
E_{\mathrm{circ}}^{(N)}(x,y)=\sum_{q=0}^{N-1}E_q(x,y),
\label{eq:steadystate_sum}
\end{equation}
and the corresponding one-way circulating power is
\begin{equation}
P_{\mathrm{circ}}^{(N)}=
\iint
\left|E_{\mathrm{circ}}^{(N)}(x,y)\right|^2\,dx\,dy.
\label{eq:power_def}
\end{equation}

All results reported in the main text were evaluated after $10^4$ round trips.
We verified numerical convergence with respect to the transverse window size,
the grid sampling, and the number of round trips included in the steady-state
reconstruction.
To quantify direct central-hole clipping, we applied the
cavity round-trip operator once to the reconstructed
circulating field obtained after $10^4$ round trips. At each mirror, the clipping fraction was
evaluated before applying the reflection operator, as the
power incident within the hole divided by the incident power
within the mirror's outer diameter.
For $D_\mathrm{h}=\quantity{100}{\text{\textmu}m}$, the losses are
approximately $10.53$ ppm at each mirror, corresponding to
a combined central-hole loss of $21.07$ ppm per round trip.

These values quantify the direct power removal
by the central-hole masks. The apertures also produce
diffraction and redistribution of the intracavity field,
which can alter its overlap with the injected mode and the
coherent cavity buildup. Outer-edge clipping, mirror
transmission, and parasitic losses further affect the
achievable circulating power. These processes are included
within our scalar Fresnel propagation model.
Figure~\ref{fig:Power vs hole diameter} therefore shows their
combined effect on the one-way circulating power after
$10^4$ round trips as the common hole diameter is varied,
with all other parameters held fixed.

For Fig.~5(b), the horizontal cut at $y=0$ was fitted to the one-dimensional
$LG_0^{1}$ intensity profile
\begin{equation}
I(x)=
A\,
\frac{2(x-x_0)^2}{w_{\mathrm{fit}}^2}
\exp\!\left[
-\frac{2(x-x_0)^2}{w_{\mathrm{fit}}^2}
\right]
+I_{\mathrm{bg}},
\label{eq:LG01_fit}
\end{equation}
where $A$ is the intensity scaling factor, $x_0$ is the lateral offset of the mode center, $w_{\mathrm{fit}}$ is the fitted beam-waist parameter at the mirror plane, and $I_{\mathrm{bg}}$ is a constant background intensity. The fitted
waist $w_{\mathrm{fit}}$ provides an effective mirror-plane radius of the
circulating $LG_0^{1}$-like mode in the presence of the central aperture.

\appendixsection{Integration into an SEM column}{app:sem_integration}

The configuration in Fig.~\ref{fig:fig2} would require a
dedicated module within the SEM condenser system. The
50~mm cavity, substrates, and mounts imply an axial space
allocation of order 0.1~m. The perforated optics would be
aligned with the electron-optical axis. Laser injection
and collection of transmitted light would use lateral
vacuum windows, with any output steering optic placed
after the second cavity mirror, outside the cavity
round-trip path. The electron-optical relay C1--C2 would
image the interaction plane onto the objective aperture
plane with $M\simeq4.2$.  Access to a
pupil-conjugate plane using additional electron optics
has been demonstrated for TEM laser phase
plates~\cite{turnbaugh2021high}.

Bore geometry, coating
integrity, optical losses, and charging would require
qualification after fabrication and assembly.

For an interaction-plane beam radius
$r_{\mathrm{IP}}\simeq6~\mu$m and semi-angle
$\theta_{\mathrm{IP}}\lesssim0.5$~mrad, the geometric
envelope at either cavity mirror satisfies
\begin{equation}
r_{\mathrm{env}}\lesssim
r_{\mathrm{IP}}+\frac{L}{2}\theta_{\mathrm{IP}}
\simeq19~\mu\mathrm{m}.
\label{eq:sem_beam_envelope}
\end{equation}
The 100~$\mu$m-diameter holes therefore provide
approximately 30~$\mu$m of radial clearance at the
reflective surfaces. Clearance through the substrates
and coupling optics would be verified separately.

At the interaction plane, $\phi_p(x,y)$ is the ponderomotive
phase. A shift $\delta x$ relative to the electron pupil gives
\begin{equation}
\begin{aligned}
\delta\phi_p &\simeq-\delta x\,\partial_x\phi_p,\\
\sigma_{\delta\phi_p} &\simeq|\delta x|
\sqrt{\left\langle(\partial_x\phi_p)^2
\right\rangle_{\mathrm{pupil}}}.
\end{aligned}
\label{eq:sem_registration_error}
\end{equation}
Here $\delta\phi_p$ is the induced phase change and
$\sigma_{\delta\phi_p}$ its rms over the uniform circular pupil.
The rms gradient along $x$ is
$\sqrt{\langle(\partial_x\phi_p)^2\rangle_{\mathrm{pupil}}}
\simeq5$~rad/$\mu$m. A 0.1~rad
additional rms budget corresponds to approximately 20~nm
of relative displacement before compensation of
displacement-induced beam tilt.

The reconstructed circulating field was propagated through
one additional passive round trip. We define the total
round-trip power-loss fraction $\mathcal{L}_{\mathrm{rt}}$
from the powers $P_{\mathrm{before}}$ and
$P_{\mathrm{after}}$, evaluated at the same reference plane
before and after this round trip:
\begin{equation}
\begin{aligned}
\mathcal{L}_{\mathrm{rt}}
&=1-\frac{P_{\mathrm{after}}}{P_{\mathrm{before}}}\\
&=1-R_1^{(\mathrm{pow})}R_2^{(\mathrm{pow})}(1-\ell_{\mathrm{mask}})
\simeq9.1123\times10^{-4}.
\end{aligned}
\label{eq:sem_roundtrip_loss}
\end{equation}
Here $R_1^{(\mathrm{pow})}=0.99921$ and $R_2^{(\mathrm{pow})}=0.99992$ include mirror
transmission and parasitic losses. The
reflectivity-corrected mask loss
$\ell_{\mathrm{mask}}=41.34$~ppm comprises
21.07~ppm from the central holes and 20.27~ppm from
outer-edge clipping. Treating this attenuation as an effective single-mode loss
gives the cavity finesse $\mathcal{F}_{\mathrm{cav}}$~\cite{wipfli2023integration}:
\begin{equation}
\mathcal{F}_{\mathrm{cav}}\simeq
\frac{2\pi}{\mathcal{L}_{\mathrm{rt}}}
\simeq6.9\times10^3.
\label{eq:sem_effective_finesse}
\end{equation}
The near-resonance circulating power, normalized to its
maximum resonant value $P_{\mathrm{circ,max}}$ for fixed
input power and coupling, is approximately~\cite{wipfli2023integration}
\begin{equation}
\frac{P_{\mathrm{circ}}}{P_{\mathrm{circ,max}}}
\simeq
\left[1+
\left(\frac{4\mathcal{F}_{\mathrm{cav}}\,
\delta L_{\mathrm{eq}}}{\lambda_L}\right)^2
\right]^{-1},
\label{eq:sem_power_detuning}
\end{equation}
where $\delta L_{\mathrm{eq}}$ denotes the equivalent
length detuning. Requiring at least 95\% of the maximum resonant power
in Eq.~\ref{eq:sem_power_detuning} gives
\begin{equation}
|\delta L_{\mathrm{eq}}|\lesssim
\frac{\lambda_L}{4\mathcal{F}_{\mathrm{cav}}}
\sqrt{\frac{1}{0.95}-1}
\simeq8.9~\mathrm{pm}.
\label{eq:sem_detuning_requirement}
\end{equation}
Using $|\delta\nu|=c|\delta L_{\mathrm{eq}}|/
(\lambda_L L)$, this corresponds to approximately
50~kHz of residual laser--cavity frequency detuning.
This requirement applies after active frequency
locking, which can track cavity-length fluctuations. 

The circulating-power sensitivity was also evaluated numerically
by varying individual parameters about the nominal cavity
configuration, while holding the input power and remaining
parameters fixed. Each calculation used $10^4$ round trips.
The resulting power was normalized to
$P_{\mathrm{circ,ref}}$, the circulating power of the nominal
configuration evaluated with the same iteration count.
Table~\ref{tab:cavity_power_tolerances} lists the sampled
perturbations closest to 95\% power retention. These values
quantify optical-power tolerances; allowable hole offsets must
additionally satisfy the electron-beam clearance condition
in Eq.~\ref{eq:sem_beam_envelope}. The analytical detuning bound is an effective single-mode
estimate based on the mask attenuation of the nominal
reconstructed field. An underestimate of the effective modal
loss would overestimate the finesse and yield an overly
restrictive detuning bound. The numerical power-response scans
in Table~\ref{tab:cavity_power_tolerances} directly evaluate
the propagated field and therefore provide the reference
values for the implemented cavity model. Both approaches
indicate residual detuning requirements on the picometre scale.

\begin{table}[htbp]
\centering
\caption{Numerical circulating-power tolerances after
$10^4$ round trips at $P_{\mathrm{in}}=40$~W.
Retained power is expressed as
$100P_{\mathrm{circ}}/P_{\mathrm{circ,ref}}$.
The focal-length change $\Delta f_a$ is relative to the
nominal anaclastic focal length $f_a=36.2$~mm, and $d_h$
denotes both holes common displacement.
For the elliptical hole, the semi-major radius $a_h$ is
varied while the semi-minor radius is fixed at the nominal
circular-hole radius, $b_h=r_h=50~\mu$m.}
\label{tab:cavity_power_tolerances}
\begin{tabular}{llc}
\hline
Perturbation & Applied change & Retained power (\%) \\
\hline
Focal-length increase
& $\Delta f_a=+75\,\mu\mathrm{m}$ & $\sim96$ \\
Cavity detuning
& $\delta L_{\mathrm{eq}}=15 \mathrm{pm}$ & $\sim94$ \\
Hole offset
& $d_h=37\,\mu\mathrm{m}$ & $\sim95$ \\
Hole ellipticity
& $a_h=70\,\mu\mathrm{m}$ & $\sim95$ \\
\hline
\end{tabular}
\end{table}

To estimate circulating-power sensitivity, we hold the
normalized optical mode and the applied defocus fixed.
Let $P_0\simeq34~\mathrm{kW}$ denote the nominal circulating
power, with the input laser power chosen to reach this value
on resonance, and $\phi_{\mathrm{p,obj}}^{(0)}$ the corresponding
ponderomotive phase in the objective-pupil plane.
We define the fractional power change as
$\epsilon=P_{\mathrm{circ}}/P_0-1$.
The quantity $\sigma_{\mathrm{add}}$ denotes the
piston-removed rms difference between the perturbed and
nominal ponderomotive phases. Here,
$\operatorname{std}_{\mathrm{pupil}}$ denotes the standard
deviation over the uniform circular pupil, calculated
after subtracting the mean pupil phase.
The power dependence is therefore
\begin{equation}
\begin{aligned}
\phi_{\mathrm{p,obj}}(\epsilon)
&=(1+\epsilon)\phi_{\mathrm{p,obj}}^{(0)},\\
\sigma_{\mathrm{add}}
&=|\epsilon|\,
\operatorname{std}_{\mathrm{pupil}}
\!\left(\phi_{\mathrm{p,obj}}^{(0)}\right)
\simeq10.713|\epsilon|~\mathrm{rad}.
\end{aligned}
\label{eq:sem_power_phase_sensitivity}
\end{equation}
The total residual rms, evaluated from the scaled correcting
phase minus the lens-aberration phase, is 0.432~rad at nominal
power, 0.505~rad for a 1\% power decrease, and 0.375~rad for
a 1\% increase. An illustrative additional phase-error budget
of 0.1~rad rms therefore requires
$|\epsilon|\lesssim0.00933$, corresponding to approximately
0.933\% circulating-power stability. At nominal resonance,
Eq.~\eqref{eq:sem_power_detuning} translates this criterion
into approximately 3.7~pm of residual equivalent length
detuning, or 21~kHz of frequency detuning. Active laser--cavity frequency locking has already been
demonstrated in TEM-integrated high-power near-concentric
laser phase plates \cite{schwartz2019laser,turnbaugh2021high}, providing an experimental
precedent for the active control of detuning.
The spatial and frequency requirements are engineering
estimates requiring verification in a prototype.

\FloatBarrier
\bibliographystyle{unsrtnat}
\bibliography{literatur}

\providecommand{\noopsort}[1]{}\providecommand{\singleletter}[1]{#1}
\begin{thebibliography}{74}
\providecommand{\natexlab}[1]{#1}
\providecommand{\url}[1]{\texttt{#1}}
\expandafter\ifx\csname urlstyle\endcsname\relax
  \providecommand{\doi}[1]{doi: #1}\else
  \providecommand{\doi}{doi: \begingroup \urlstyle{rm}\Url}\fi

\bibitem[Jobst et~al.(2016)Jobst, Van~der Torren, Krasovskii, Balgley, Dean, Tromp, and Van~der Molen]{jobst2016quantifying}
Johannes Jobst, Alexander~JH Van~der Torren, Eugene~E Krasovskii, Jesse Balgley, Cory~R Dean, Rudolf~M Tromp, and Sense~Jan Van~der Molen.
\newblock Quantifying electronic band interactions in van der waals materials using angle-resolved reflected-electron spectroscopy.
\newblock \emph{Nature Communications}, 7\penalty0 (1):\penalty0 13621, 2016.

\bibitem[Geelen et~al.(2019)Geelen, Jobst, Krasovskii, van~der Molen, and Tromp]{geelen2019nonuniversal}
Dani{\"e}l Geelen, J~Jobst, EE~Krasovskii, Sense~Jan van~der Molen, and Rudolf~M Tromp.
\newblock Nonuniversal transverse electron mean free path through few-layer graphene.
\newblock \emph{Physical review letters}, 123\penalty0 (8):\penalty0 086802, 2019.

\bibitem[Sunaoshi et~al.(2016)Sunaoshi, Kaji, Orai, Schamp, and Voelkl]{sunaoshi2016stem}
Takeshi Sunaoshi, Kazutoshi Kaji, Yoshihisa Orai, CT~Schamp, and Edgar Voelkl.
\newblock Stem/sem, chemical analysis, atomic resolution and surface imaging at {$\leq 30$}~{kV} with no aberration correction for nanomaterials on graphene support.
\newblock \emph{Microscopy and Microanalysis}, 22\penalty0 (S3):\penalty0 604--605, 2016.

\bibitem[Chirita et~al.(2022)Chirita, Markevich, Tripathi, Pike, Verstraete, Kotakoski, and Susi]{chirita2022three}
Alexandru Chirita, Alexander Markevich, Mukesh Tripathi, Nicholas~A Pike, Matthieu~J Verstraete, Jani Kotakoski, and Toma Susi.
\newblock Three-dimensional ab initio description of vibration-assisted electron knock-on displacements in graphene.
\newblock \emph{Physical Review B}, 105\penalty0 (23):\penalty0 235419, 2022.

\bibitem[Shiloh et~al.(2022)Shiloh, Chlouba, and Hommelhoff]{shiloh2022quantum}
R~Shiloh, T~Chlouba, and P~Hommelhoff.
\newblock Quantum-coherent light-electron interaction in a scanning electron microscope.
\newblock \emph{Physical Review Letters}, 128\penalty0 (23):\penalty0 235301, 2022.

\bibitem[Simonaitis et~al.(2025)Simonaitis, Alongi, Slayton, Putnam, Berggren, and Keathley]{simonaitis2025electron}
John~W Simonaitis, Joseph~A Alongi, Benjamin Slayton, William~P Putnam, Karl~K Berggren, and Phillip~D Keathley.
\newblock Electron energy loss spectroscopy of two-dimensional materials in a scanning electron microscope.
\newblock \emph{Physical Review B}, 112\penalty0 (23):\penalty0 235421, 2025.

\bibitem[Lagos et~al.(2022)Lagos, Bicket, Mousavi~M, and Botton]{lagos2022advances}
Maureen~J Lagos, Isobel~C Bicket, S~Shayan Mousavi~M, and Gianluigi~A Botton.
\newblock Advances in ultrahigh-energy resolution eels: phonons, infrared plasmons and strongly coupled modes.
\newblock \emph{Microscopy}, 71\penalty0 (Supplement\_1):\penalty0 i174--i199, 2022.

\bibitem[Verbeeck et~al.(2010)Verbeeck, Tian, and Schattschneider]{verbeeck2010production}
Johan Verbeeck, He~Tian, and Peter Schattschneider.
\newblock Production and application of electron vortex beams.
\newblock \emph{Nature}, 467\penalty0 (7313):\penalty0 301--304, 2010.

\bibitem[McMorran et~al.(2011)McMorran, Agrawal, Anderson, Herzing, Lezec, McClelland, and Unguris]{mcmorran2011electron}
Benjamin~J McMorran, Amit Agrawal, Ian~M Anderson, Andrew~A Herzing, Henri~J Lezec, Jabez~J McClelland, and John Unguris.
\newblock Electron vortex beams with high quanta of orbital angular momentum.
\newblock \emph{science}, 331\penalty0 (6014):\penalty0 192--195, 2011.

\bibitem[Grillo et~al.(2014)Grillo, Karimi, Gazzadi, Frabboni, Dennis, and Boyd]{grillo2014}
Vincenzo Grillo, Ebrahim Karimi, Gian~Carlo Gazzadi, Stefano Frabboni, Mark~R. Dennis, and Robert~W. Boyd.
\newblock Generation of nondiffracting electron bessel beams.
\newblock \emph{Phys. Rev. X}, 4:\penalty0 011013, Jan 2014.
\newblock \doi{10.1103/PhysRevX.4.011013}.

\bibitem[Haider et~al.(1998)Haider, Uhlemann, Schwan, Rose, Kabius, and Urban]{haider1998electron}
Maximilian Haider, Stephan Uhlemann, Eugen Schwan, Harald Rose, Bernd Kabius, and Knut Urban.
\newblock Electron microscopy image enhanced.
\newblock \emph{Nature}, 392\penalty0 (6678):\penalty0 768--769, 1998.

\bibitem[Nakano and Yamazawa(2026)]{nakano2025development}
Tomonori Nakano and Yu~Yamazawa.
\newblock Development of a wire corrector for low accelerating voltage scanning electron microscopy.
\newblock \emph{Microscopy}, 75\penalty0 (Supplement 1):\penalty0 i52--i59, 2026.
\newblock \doi{10.1093/jmicro/dfaf054}.

\bibitem[Barwick et~al.(2009)Barwick, Flannigan, and Zewail]{barwick2009photon}
Brett Barwick, David~J Flannigan, and Ahmed~H Zewail.
\newblock Photon-induced near-field electron microscopy.
\newblock \emph{Nature}, 462\penalty0 (7275):\penalty0 902--906, 2009.

\bibitem[Feist et~al.(2015)Feist, Echternkamp, Schauss, Yalunin, Sch{\"a}fer, and Ropers]{feist2015quantum}
Armin Feist, Katharina~E Echternkamp, Jakob Schauss, Sergey~V Yalunin, Sascha Sch{\"a}fer, and Claus Ropers.
\newblock Quantum coherent optical phase modulation in an ultrafast transmission electron microscope.
\newblock \emph{Nature}, 521\penalty0 (7551):\penalty0 200--203, 2015.

\bibitem[Vanacore et~al.(2019)Vanacore, Berruto, Madan, Pomarico, Biagioni, Lamb, McGrouther, Reinhardt, Kaminer, Barwick, et~al.]{vanacore2019ultrafast}
Giovanni~Maria Vanacore, Gabriele Berruto, Ivan Madan, Enrico Pomarico, P~Biagioni, RJ~Lamb, D~McGrouther, O~Reinhardt, I~Kaminer, B~Barwick, et~al.
\newblock Ultrafast generation and control of an electron vortex beam via chiral plasmonic near fields.
\newblock \emph{Nature materials}, 18\penalty0 (6):\penalty0 573--579, 2019.

\bibitem[Kone{\v{c}}n{\'a} and de~Abajo(2020)]{konevcna2020electron}
Andrea Kone{\v{c}}n{\'a} and F~Javier~Garc{\'\i}a de~Abajo.
\newblock Electron beam aberration correction using optical near fields.
\newblock \emph{Physical Review Letters}, 125\penalty0 (3):\penalty0 030801, 2020.

\bibitem[Shiloh et~al.(2021)Shiloh, Illmer, Chlouba, Yousefi, Sch{\"o}nenberger, Niedermayer, Mittelbach, and Hommelhoff]{shiloh2021electron}
Roy Shiloh, Johannes Illmer, Tom{\'a}{\v{s}} Chlouba, Peyman Yousefi, Norbert Sch{\"o}nenberger, Uwe Niedermayer, Anna Mittelbach, and Peter Hommelhoff.
\newblock Electron phase-space control in photonic chip-based particle acceleration.
\newblock \emph{Nature}, 597\penalty0 (7877):\penalty0 498--502, 2021.

\bibitem[Henke et~al.(2021)Henke, Raja, Feist, Huang, Arend, Yang, Kappert, Wang, M{\"o}ller, Pan, et~al.]{henke2021integrated}
Jan-Wilke Henke, Arslan~Sajid Raja, Armin Feist, Guanhao Huang, Germaine Arend, Yujia Yang, F~Jasmin Kappert, Rui~Ning Wang, Marcel M{\"o}ller, Jiahe Pan, et~al.
\newblock Integrated photonics enables continuous-beam electron phase modulation.
\newblock \emph{Nature}, 600\penalty0 (7890):\penalty0 653--658, 2021.

\bibitem[Feist et~al.(2022)Feist, Huang, Arend, Yang, Henke, Raja, Kappert, Wang, Louren{\c{c}}o-Martins, Qiu, et~al.]{feist2022cavity}
Armin Feist, Guanhao Huang, Germaine Arend, Yujia Yang, Jan-Wilke Henke, Arslan~Sajid Raja, F~Jasmin Kappert, Rui~Ning Wang, Hugo Louren{\c{c}}o-Martins, Zheru Qiu, et~al.
\newblock Cavity-mediated electron-photon pairs.
\newblock \emph{Science}, 377\penalty0 (6607):\penalty0 777--780, 2022.

\bibitem[Madan et~al.(2022)Madan, Leccese, Mazur, Barantani, LaGrange, Sapozhnik, Tengdin, Gargiulo, Rotunno, Olaya, et~al.]{madan2022ultrafast}
Ivan Madan, Veronica Leccese, Adam Mazur, Francesco Barantani, Thomas LaGrange, Alexey Sapozhnik, Phoebe~M Tengdin, Simone Gargiulo, Enzo Rotunno, Jean-Christophe Olaya, et~al.
\newblock Ultrafast transverse modulation of free electrons by interaction with shaped optical fields.
\newblock \emph{ACS photonics}, 9\penalty0 (10):\penalty0 3215--3224, 2022.

\bibitem[Tsesses et~al.(2023)Tsesses, Dahan, Wang, Bucher, Cohen, Reinhardt, Bartal, and Kaminer]{tsesses2023tunable}
Shai Tsesses, Raphael Dahan, Kangpeng Wang, Tomer Bucher, Kobi Cohen, Ori Reinhardt, Guy Bartal, and Ido Kaminer.
\newblock Tunable photon-induced spatial modulation of free electrons.
\newblock \emph{Nature Materials}, 22\penalty0 (3):\penalty0 345--352, 2023.

\bibitem[Garc{\'\i}a~de Abajo and Ropers(2023)]{garcia2023spatiotemporal}
F~Javier Garc{\'\i}a~de Abajo and Claus Ropers.
\newblock Spatiotemporal electron beam focusing through parallel interactions with shaped optical fields.
\newblock \emph{Physical Review Letters}, 130\penalty0 (24):\penalty0 246901, 2023.

\bibitem[Fang et~al.(2024)Fang, Kuttruff, Nabben, and Baum]{fang2024structured}
Yiqi Fang, Joel Kuttruff, David Nabben, and Peter Baum.
\newblock Structured electrons with chiral mass and charge.
\newblock \emph{Science}, 385\penalty0 (6705):\penalty0 183--187, 2024.

\bibitem[Ferrari et~al.(2025)Ferrari, Duncan, Yannai, Dahan, Rosi, Ostroman, Bravi, Niedermayr, Abudi, Adiv, et~al.]{ferrari2025realization}
Beatrice~Matilde Ferrari, Cameron James~Richard Duncan, Michael Yannai, Raphael Dahan, Paolo Rosi, Irene Ostroman, Maria~Giulia Bravi, Arthur Niedermayr, Tom~Lenkiewicz Abudi, Yuval Adiv, et~al.
\newblock Realization of a pre-sample photonic-based free-electron modulator in ultrafast transmission electron microscopes.
\newblock \emph{ACS photonics}, 12\penalty0 (11):\penalty0 5864--5873, 2025.

\bibitem[Garc{\'\i}a~de Abajo et~al.(2025)Garc{\'\i}a~de Abajo, Polman, Velasco, Kociak, Tizei, St{\'e}phan, Meuret, Sannomiya, Akiba, Auad, et~al.]{garcia2025roadmap}
F~Javier Garc{\'\i}a~de Abajo, Albert Polman, Cruz~I Velasco, Mathieu Kociak, Luiz~HG Tizei, Odile St{\'e}phan, Sophie Meuret, Takumi Sannomiya, Keiichirou Akiba, Yves Auad, et~al.
\newblock Roadmap for quantum nanophotonics with free electrons.
\newblock \emph{ACS photonics}, 12\penalty0 (9):\penalty0 4760--4817, 2025.

\bibitem[Chahshouri et~al.(2026)Chahshouri, Ebel, Funk, and Talebi]{chahshouri2026stimulated}
Fatemeh Chahshouri, Sven Ebel, Mitja Funk, and Nahid Talebi.
\newblock Stimulated interactions of low-energy free-electrons with light.
\newblock \emph{Advances in Physics: X}, 11\penalty0 (1):\penalty0 2647966, 2026.

\bibitem[Shiloh et~al.(2018)Shiloh, Remez, Lu, Jin, Lereah, Tavabi, Dunin-Borkowski, and Arie]{shiloh2018spherical}
Roy Shiloh, Roei Remez, Peng-Han Lu, Lei Jin, Yossi Lereah, Amir~H Tavabi, Rafal~E Dunin-Borkowski, and Ady Arie.
\newblock Spherical aberration correction in a scanning transmission electron microscope using a sculpted thin film.
\newblock \emph{Ultramicroscopy}, 189:\penalty0 46--53, 2018.

\bibitem[Tavabi et~al.(2021)Tavabi, Rosi, Rotunno, Roncaglia, Belsito, Frabboni, Pozzi, Gazzadi, Lu, Nijland, Ghosh, Tiemeijer, Karimi, Dunin-Borkowski, and Grillo]{Tavabi2021}
Amir~H. Tavabi, Paolo Rosi, Enzo Rotunno, Alberto Roncaglia, Luca Belsito, Stefano Frabboni, Giulio Pozzi, Gian~Carlo Gazzadi, Peng-Han Lu, Robert Nijland, Moumita Ghosh, Peter Tiemeijer, Ebrahim Karimi, Rafal~E. Dunin-Borkowski, and Vincenzo Grillo.
\newblock Experimental demonstration of an electrostatic orbital angular momentum sorter for electron beams.
\newblock \emph{Phys. Rev. Lett.}, 126:\penalty0 094802, Mar 2021.
\newblock \doi{10.1103/PhysRevLett.126.094802}.

\bibitem[Vega~Ib{\'a}{\~n}ez et~al.(2023)Vega~Ib{\'a}{\~n}ez, B{\'e}ch{\'e}, and Verbeeck]{vega2023can}
Francisco Vega~Ib{\'a}{\~n}ez, Armand B{\'e}ch{\'e}, and Johan Verbeeck.
\newblock Can a programmable phase plate serve as an aberration corrector in the transmission electron microscope (tem)?
\newblock \emph{Microscopy and Microanalysis}, 29\penalty0 (1):\penalty0 341--351, 2023.

\bibitem[Adelmark et~al.(2026)Adelmark, Kavkani, Chalangar, Taboryski, Lavrinenko, Beleggia, and Bunea]{adelmark2026fabrication}
Mathias~V Adelmark, Payam~H Kavkani, Ebrahim Chalangar, Rafael Taboryski, Andrei Lavrinenko, Marco Beleggia, and Ada-Ioana Bunea.
\newblock Fabrication and characterization of an adaptive thin-film electrostatic phase plate.
\newblock \emph{Physical Review Applied}, 25\penalty0 (5):\penalty0 054007, 2026.

\bibitem[Freimund et~al.(2001)Freimund, Aflatooni, and Batelaan]{freimund2001observation}
Daniel~L Freimund, Kayvan Aflatooni, and Herman Batelaan.
\newblock Observation of the kapitza--dirac effect.
\newblock \emph{Nature}, 413\penalty0 (6852):\penalty0 142--143, 2001.

\bibitem[Hebeisen et~al.(2006)Hebeisen, Ernstorfer, Harb, Dartigalongue, Jordan, and Miller]{hebeisen2006femtosecond}
Christoph~T Hebeisen, Ralph Ernstorfer, Maher Harb, Thibault Dartigalongue, Robert~E Jordan, and RJ~Dwayne Miller.
\newblock Femtosecond electron pulse characterization using laser ponderomotive scattering.
\newblock \emph{Optics letters}, 31\penalty0 (23):\penalty0 3517--3519, 2006.

\bibitem[de~Abajo and Kone{\v{c}}n{\'a}(2021)]{de2021optical}
F~Javier~Garc{\'\i}a de~Abajo and Andrea Kone{\v{c}}n{\'a}.
\newblock Optical modulation of electron beams in free space.
\newblock \emph{Physical Review Letters}, 126\penalty0 (12):\penalty0 123901, 2021.

\bibitem[Chirita~Mihaila et~al.(2022)Chirita~Mihaila, Weber, Schneller, Grandits, Nimmrichter, and Juffmann]{chirita2022transverse}
Marius~Constantin Chirita~Mihaila, Philipp Weber, Matthias Schneller, Lucas Grandits, Stefan Nimmrichter, and Thomas Juffmann.
\newblock Transverse electron-beam shaping with light.
\newblock \emph{Physical Review X}, 12\penalty0 (3):\penalty0 031043, 2022.

\bibitem[Chirita~Mihaila and Koz{\'a}k(2025)]{chirita2025design}
Marius~Constantin Chirita~Mihaila and Martin Koz{\'a}k.
\newblock Design for light-based spherical aberration correction of ultrafast electron microscopes.
\newblock \emph{Optics Express}, 33\penalty0 (1):\penalty0 758--775, 2025.

\bibitem[Mihaila et~al.(2025)Mihaila, La{\v{s}}tovi{\v{c}}kov{\'a}~Streshkova, and Koz{\'a}k]{mihaila2025light}
Marius Constantin~Chirita Mihaila, Neli La{\v{s}}tovi{\v{c}}kov{\'a}~Streshkova, and Martin Koz{\'a}k.
\newblock Light-based chromatic aberration correction of ultrafast electron microscopes.
\newblock \emph{Physical Review Letters}, 134\penalty0 (20):\penalty0 203802, 2025.

\bibitem[Chirita~Mihaila et~al.(2025)Chirita~Mihaila, Koutensk{\`y}, Moriov{\'a}, and Koz{\'a}k]{chirita2025light}
Marius~Constantin Chirita~Mihaila, Petr Koutensk{\`y}, Kamila Moriov{\'a}, and Martin Koz{\'a}k.
\newblock Light-based electron aberration corrector.
\newblock \emph{Nature Photonics}, 19:\penalty0 1309–1314, 2025.

\bibitem[Uesugi et~al.(2021)Uesugi, Kozawa, and Sato]{uesugi2021electron}
Yuuki Uesugi, Yuichi Kozawa, and Shunichi Sato.
\newblock Electron round lenses with negative spherical aberration by a tightly focused cylindrically polarized light beam.
\newblock \emph{Physical Review Applied}, 16\penalty0 (1):\penalty0 L011002, 2021.

\bibitem[Streshkova et~al.(2024)Streshkova, Koutensk{\`y}, Novotn{\`y}, and Koz{\'a}k]{streshkova2024monochromatization}
Neli~La{\v{s}}tovi{\v{c}}kov{\'a} Streshkova, Petr Koutensk{\`y}, Tom{\'a}{\v{s}} Novotn{\`y}, and Martin Koz{\'a}k.
\newblock Monochromatization of electron beams with spatially and temporally modulated optical fields.
\newblock \emph{Physical Review Letters}, 133\penalty0 (21):\penalty0 213801, 2024.

\bibitem[Uesugi et~al.(2022)Uesugi, Kozawa, and Sato]{uesugi2022properties}
Yuuki Uesugi, Yuichi Kozawa, and Shunichi Sato.
\newblock Properties of electron lenses produced by ponderomotive potential with bessel and laguerre--gaussian beams.
\newblock \emph{Journal of Optics}, 24\penalty0 (5):\penalty0 054013, 2022.

\bibitem[Uesugi and Kozawa(2025)]{uesugi2025crossed}
Yuuki Uesugi and Yuichi Kozawa.
\newblock Crossed ponderomotive lenses for spherical aberration correction in electron optics.
\newblock \emph{Physical Review A}, 112\penalty0 (1):\penalty0 013507, 2025.

\bibitem[Nekula et~al.(2025)Nekula, Juffmann, and Kone{\v{c}}n{\'a}]{nekula2025compensating}
Zden{\v{e}}k Nekula, Thomas Juffmann, and Andrea Kone{\v{c}}n{\'a}.
\newblock Compensating spherical and chromatic aberrations of ultrafast electron microscopes with laser beams.
\newblock \emph{Physical Review Research}, 7\penalty0 (4):\penalty0 043024, 2025.

\bibitem[Uesugi and Kozawa(2026)]{uesugi2026achromat}
Yuuki Uesugi and Yuichi Kozawa.
\newblock Ponderomotive achromat for electron optics: Radially polarized annular focusing and a round-lens corrector regime.
\newblock \emph{Physical Review Research}, 8\penalty0 (3):\penalty0 033300, 2026.
\newblock \doi{10.1103/5s6w-2dt5}.

\bibitem[Ebel and Talebi(2025)]{ebel2025structured}
Sven Ebel and Nahid Talebi.
\newblock Structured free-space optical fields for transverse and longitudinal control of electron matter waves.
\newblock \emph{New Journal of Physics}, 27\penalty0 (5):\penalty0 054103, 2025.

\bibitem[Velasco and Garc{\'\i}a~de Abajo(2025)]{velasco2025free}
Cruz~I Velasco and F~Javier Garc{\'\i}a~de Abajo.
\newblock Free-space optical modulation of free electrons in the continuous-wave regime.
\newblock \emph{Physical Review Letters}, 134\penalty0 (12):\penalty0 123804, 2025.

\bibitem[Zhao et~al.(2025)Zhao, Fang, Uluda{\u{g}}, and Hommelhoff]{zhao2025gouy}
Zhexin Zhao, Yiqi Fang, Mevlana~Yunus Uluda{\u{g}}, and Peter Hommelhoff.
\newblock Gouy phase-related effects in the free-space optical modulation of free electrons.
\newblock \emph{arXiv preprint arXiv:2510.24939}, 2025.

\bibitem[Zheng and Kfir(2026)]{zheng2026elastic}
Dingguo Zheng and Ofer Kfir.
\newblock Elastic quantum coupling between free electrons and photons.
\newblock \emph{Science Advances}, 12:\penalty0 eaed3278, 2026.
\newblock \doi{10.1126/sciadv.aed3278}.

\bibitem[Durak et~al.(2014)Durak, Nguyen, Leong, Straupe, and Kurtsiefer]{durak2014diffraction}
Kadir Durak, Chi~Huan Nguyen, Victor Leong, Stanislav Straupe, and Christian Kurtsiefer.
\newblock Diffraction-limited fabry--perot cavity in the near concentric regime.
\newblock \emph{New Journal of Physics}, 16\penalty0 (10):\penalty0 103002, 2014.

\bibitem[Nguyen et~al.(2018)Nguyen, Utama, Lewty, and Kurtsiefer]{nguyen2018operating}
Chi~Huan Nguyen, Adrian~Nugraha Utama, Nick Lewty, and Christian Kurtsiefer.
\newblock Operating a near-concentric cavity at the last stable resonance.
\newblock \emph{Physical Review A}, 98\penalty0 (6):\penalty0 063833, 2018.

\bibitem[Utama et~al.(2021)Utama, Chow, Nguyen, and Kurtsiefer]{utama2021coupling}
Adrian~Nugraha Utama, Chang~Hoong Chow, Chi~Huan Nguyen, and Christian Kurtsiefer.
\newblock Coupling light to higher order transverse modes of a near-concentric optical cavity.
\newblock \emph{Optics Express}, 29\penalty0 (6):\penalty0 8130--8141, 2021.

\bibitem[Schwartz et~al.(2017)Schwartz, Axelrod, Tuthill, Haslinger, Ophus, Glaeser, and M{\"u}ller]{schwartz2017near}
Osip Schwartz, JJ~Axelrod, DR~Tuthill, P~Haslinger, C~Ophus, RM~Glaeser, and H~M{\"u}ller.
\newblock Near-concentric fabry-p{\'e}rot cavity for continuous-wave laser control of electron waves.
\newblock \emph{Optics express}, 25\penalty0 (13):\penalty0 14453--14462, 2017.

\bibitem[Schwartz et~al.(2019)Schwartz, Axelrod, Campbell, Turnbaugh, Glaeser, and M{\"u}ller]{schwartz2019laser}
Osip Schwartz, Jeremy~J Axelrod, Sara~L Campbell, Carter Turnbaugh, Robert~M Glaeser, and Holger M{\"u}ller.
\newblock Laser phase plate for transmission electron microscopy.
\newblock \emph{Nature methods}, 16\penalty0 (10):\penalty0 1016--1020, 2019.

\bibitem[Axelrod et~al.(2020)Axelrod, Campbell, Schwartz, Turnbaugh, Glaeser, and M\"uller]{PhysRevLett.124.174801}
Jeremy~J. Axelrod, Sara~L. Campbell, Osip Schwartz, Carter Turnbaugh, Robert~M. Glaeser, and Holger M\"uller.
\newblock Observation of the relativistic reversal of the ponderomotive potential.
\newblock \emph{Phys. Rev. Lett.}, 124:\penalty0 174801, May 2020.
\newblock \doi{10.1103/PhysRevLett.124.174801}.

\bibitem[Petrov et~al.(2026{\natexlab{a}})Petrov, Zhang, Axelrod, Olshin, and M{\"u}ller]{petrov2026crossed}
Petar~N. Petrov, Jessie~T. Zhang, Jeremy~J. Axelrod, Pavel~K. Olshin, and Holger M{\"u}ller.
\newblock Crossed laser phase plates for transmission electron microscopy.
\newblock \emph{Nature Communications}, 17:\penalty0 7199, June 2026{\natexlab{a}}.
\newblock \doi{10.1038/s41467-026-74060-6}.

\bibitem[Petrov et~al.(2026{\natexlab{b}})Petrov, Zhang, Remis, Axelrod, Cheng, Cooper, Hicklin, Sandhaus, Schnurr, Glaeser, and M{\"u}ller]{Petrov2026LaserPhasePlate}
Petar~N. Petrov, Jessie~T. Zhang, Jonathan Remis, Jeremy~J. Axelrod, Hang Cheng, Eric~S. Cooper, Ian~K. Hicklin, Shahar Sandhaus, Cooper Schnurr, Robert~M. Glaeser, and Holger M{\"u}ller.
\newblock Laser phase plate improves structure determination of small proteins by cryo-{EM}.
\newblock \emph{Science}, 393\penalty0 (6807):\penalty0 195--199, July 2026{\natexlab{b}}.
\newblock \doi{10.1126/science.aeh0665}.

\bibitem[Arakcheev et~al.(2026)Arakcheev, Barkai, Vasilyev, and Schwartz]{arakcheev2026probing}
Alexey Arakcheev, Niv Barkai, Alexandr Vasilyev, and Osip Schwartz.
\newblock Probing the electromagnetic nonlinearity of vacuum with continuous-wave lasers.
\newblock \emph{arXiv preprint arXiv:2605.08359}, 2026.

\bibitem[Turnbaugh et~al.(2021)Turnbaugh, Axelrod, Campbell, Dioquino, Petrov, Remis, Schwartz, Yu, Cheng, Glaeser, and Mueller]{turnbaugh2021high}
Carter Turnbaugh, Jeremy~J. Axelrod, Sara~L. Campbell, Jeske~Y. Dioquino, Petar~N. Petrov, Jonathan Remis, Osip Schwartz, Zanlin Yu, Yifan Cheng, Robert~M. Glaeser, and Holger Mueller.
\newblock High-power near-concentric {Fabry--Perot} cavity for phase contrast electron microscopy.
\newblock \emph{Review of Scientific Instruments}, 92\penalty0 (5):\penalty0 053005, 2021.
\newblock \doi{10.1063/5.0045496}.

\bibitem[Zhu and Wang(2014)]{zhu2014arbitrary}
Long Zhu and Jian Wang.
\newblock Arbitrary manipulation of spatial amplitude and phase using phase-only spatial light modulators.
\newblock \emph{Scientific reports}, 4\penalty0 (1):\penalty0 7441, 2014.

\bibitem[Beresna et~al.(2011)Beresna, Gecevi{\v{c}}ius, Kazansky, and Gertus]{beresna2011radially}
Martynas Beresna, Mindaugas Gecevi{\v{c}}ius, Peter~G. Kazansky, and Titas Gertus.
\newblock Radially polarized optical vortex converter created by femtosecond laser nanostructuring of glass.
\newblock \emph{Applied Physics Letters}, 98\penalty0 (20):\penalty0 201101, 2011.
\newblock \doi{10.1063/1.3590716}.

\bibitem[Fienup(2024)]{fienup2024fourier}
James~R Fienup.
\newblock Fourier-optics imaging analysis with abcd matrices: tutorial.
\newblock \emph{Journal of the Optical Society of America A}, 41\penalty0 (12):\penalty0 2361--2370, 2024.

\bibitem[Scherzer(1947)]{scherzer1947spharische}
O~Scherzer.
\newblock Spharische und chromatische korrektur von elektronen-linsen.
\newblock \emph{Optik}, 2:\penalty0 114--132, 1947.

\bibitem[Haider et~al.(2000)Haider, Uhlemann, and Zach]{haider2000upper}
M~Haider, S~Uhlemann, and J~Zach.
\newblock Upper limits for the residual aberrations of a high-resolution aberration-corrected stem.
\newblock \emph{Ultramicroscopy}, 81\penalty0 (3-4):\penalty0 163--175, 2000.

\bibitem[Adriaans et~al.(2026)Adriaans, Hoogenboom, and Mohammadi-Gheidari]{ADRIAANS2026114297}
M.J. Adriaans, J.P. Hoogenboom, and A.~Mohammadi-Gheidari.
\newblock Basic considerations in the design of an electrostatic electron monochromator.
\newblock \emph{Ultramicroscopy}, 281:\penalty0 114297, 2026.
\newblock \doi{https://doi.org/10.1016/j.ultramic.2025.114297}.

\bibitem[Lembessis et~al.(2024)Lembessis, Koksal, Babiker, and Yuan]{lembessis2024miniature}
Vasileios~E Lembessis, Koray Koksal, Mohamed Babiker, and Jun Yuan.
\newblock Miniature atom bottle traps enabled by chiral doughnut light.
\newblock \emph{Optics Express}, 32\penalty0 (8):\penalty0 13450--13466, 2024.

\bibitem[Degallaix(2020)]{degallaix2020oscar}
J{\'e}r{\^o}me Degallaix.
\newblock Oscar: A matlab based package to simulate realistic optical cavities.
\newblock \emph{SoftwareX}, 12:\penalty0 100587, 2020.

\bibitem[Kogelnik and Li(1966)]{kogelnik1966laser}
Herwig Kogelnik and Tingye Li.
\newblock Laser beams and resonators.
\newblock \emph{Applied optics}, 5\penalty0 (10):\penalty0 1550--1567, 1966.

\bibitem[Axelrod et~al.(2026)Axelrod, Maisenbacher, Singh, Pope, Petrov, Zhang, and M{\"u}ller]{axelrod2026thermoelastic}
Jeremy~J. Axelrod, Lothar Maisenbacher, Ashwin Singh, Isaac~M. Pope, Petar~N. Petrov, Jessie~T. Zhang, and Holger M{\"u}ller.
\newblock A thermoelastic limit on the focal intensity in {Fabry--P{\'e}rot} cavities, 2026.
\newblock arXiv:2604.07676.

\bibitem[Haider et~al.(2009)Haider, Hartel, M{\"u}ller, Uhlemann, and Zach]{haider2009current}
M~Haider, P~Hartel, H~M{\"u}ller, S~Uhlemann, and J~Zach.
\newblock Current and future aberration correctors for the improvement of resolution in electron microscopy.
\newblock \emph{Philosophical Transactions of the Royal Society A: Mathematical, Physical and Engineering Sciences}, 367\penalty0 (1903):\penalty0 3665--3682, 2009.

\bibitem[Uhlemann et~al.(2013)Uhlemann, M{\"u}ller, Hartel, Zach, and Haider]{uhlemann2013thermal}
Stephan Uhlemann, Heiko M{\"u}ller, Peter Hartel, Joachim Zach, and Max Haider.
\newblock Thermal magnetic field noise limits resolution in transmission electron microscopy.
\newblock \emph{Physical review letters}, 111\penalty0 (4):\penalty0 046101, 2013.

\bibitem[Ixquiac~M{\'e}ndez et~al.(2025)Ixquiac~M{\'e}ndez, Zanetti, Kraeft, and Juffmann]{ixquiac2025low}
Luis~Alfredo Ixquiac~M{\'e}ndez, Martino Zanetti, Tilman Kraeft, and Thomas Juffmann.
\newblock Low-energy single-electron detector with submicron resolution.
\newblock \emph{ACS photonics}, 13\penalty0 (1):\penalty0 298--302, 2025.

\bibitem[Chirita~Mihaila and Koz{\'a}k(2026)]{ChiritaMihaila2026ZenodoCavity}
Marius~Constantin Chirita~Mihaila and Martin Koz{\'a}k.
\newblock {Data for "Transverse Modulation of Continuous Electron Beams by a Structured Optical Cavity"}, 2026.
\newblock URL \url{https://doi.org/10.5281/zenodo.20689393}.

\bibitem[Richards and Wolf(1959)]{richards1959electromagnetic}
Bernard Richards and Emil Wolf.
\newblock Electromagnetic diffraction in optical systems, ii. structure of the image field in an aplanatic system.
\newblock \emph{Proceedings of the Royal Society of London. Series A. Mathematical and Physical Sciences}, 253\penalty0 (1274):\penalty0 358--379, 1959.

\bibitem[Dorn et~al.(2003)Dorn, Quabis, and Leuchs]{dorn2003sharper}
Ralf Dorn, Susanne Quabis, and Gerd Leuchs.
\newblock Sharper focus for a radially polarized light beam.
\newblock \emph{Physical review letters}, 91\penalty0 (23):\penalty0 233901, 2003.

\bibitem[Wipfli et~al.(2023)Wipfli, Passagem, Fischer, Grau, and Home]{wipfli2023integration}
Oliver Wipfli, Henry~Fernandes Passagem, Christoph Fischer, Matt Grau, and Jonathan~P. Home.
\newblock Integration of a high finesse cryogenic build-up cavity with an ion trap.
\newblock \emph{Review of Scientific Instruments}, 94\penalty0 (8):\penalty0 083204, 2023.
\newblock \doi{10.1063/5.0155418}.

\end{thebibliography}

\end{document}